\documentclass[apl,twocolumn,showpacs,preprintnumbers,amsmath,amssymb,superscriptaddress]{revtex4}
\usepackage{amssymb}
\usepackage{color}

\def \Prague{Institute of Physics, Charles University in Prague, Faculty of Mathematics and Physics, Ke Karlovu 5, Prague 2, CZ-121~16, Czech Republic}

\usepackage{epsfig}
\usepackage{amsmath}
\begin{document}
\title{Effect of residual gas composition on epitaxial graphene growth on SiC}
\author{J. \surname{Kunc}} \affiliation{\Prague}
\email{kunc@karlov.mff.cuni.cz}
\author{M. \surname{Rejhon}}
\author{E. \surname{Belas}} \affiliation{\Prague}
\author{V. \surname{D\v{e}di\v{c}}} \affiliation{\Prague}
\author{P. \surname{Moravec}} \affiliation{\Prague}
\author{J. \surname{Franc}} \affiliation{\Prague}

\date{\today}

\begin{abstract}
In recent years, graphene growth optimization has been one of the key routes towards large-scale, high-quality graphene production. We have measured in-situ residual gas content during epitaxial graphene growth on silicon carbide (SiC) to find detrimental factors of epitaxial graphene growth. The growth conditions in high vacuum and purified argon are compared. The grown epitaxial graphene is studied by Raman scattering mapping and mechanical strain, charge density, number of graphene layers and graphene grain size are evaluated. Charge density and carrier mobility has been studied by Hall effect measurements in van der Pauw configuration. 
We have identified a major role of chemical reaction of carbon and residual water. The rate of the reaction is lowered when purified argon is used. We also show, that according to time varying gas content, it is preferable to grow graphene at higher temperatures and shorter times. Other sources of growth environment contamination are also discussed. The reaction of water and carbon is discussed to be one of the factors increasing number of defects in graphene. The importance of purified argon and its sufficient flow rate is concluded to be important for high-quality graphene growth as it reduces the rate of undesired chemical reactions and provides more stable and defined growth ambient. 
\end{abstract}

%\pacs{78.67.-n, 73.21.-b}

\maketitle
\section{Introduction}

Epitaxial graphene growth on SiC~\cite{Berger2004} is a scalable fabrication method of high quality graphene for post-silicon electronics and opto-electronics~\cite{KuncNanoLetters2014,RuiJPDApplPhys2014}. Among novel devices, the growth technique, together with SiC wafer preparation, stands at the beginning of whole manufacturing process. It is therefore a key to understand conditions under which reproducible high quality graphene can be reached~\cite{YazdiCrystals2016,NemecPRB2015}. 
%%%%%%Interest in improved graphene growth%%%%%%%%%%%%%%%%%%%%%%%%%%%%%%%%%%%%%%%%%%%%%%%%%%%%%%%%%%%%%%%%%%%%%%%%%%%%%%%%%%%%%%%%%%%%%
The growth mechanisms has been studied both theoretically and experimentally. First-principles calculations has been done to study graphene buffer layer formation on SiC(0001), diffusion of carbon on SiC~\cite{InouePRB2012}, stability and reactivity of atomic steps of SiC in the initial graphene growth stage~\cite{KageshimaPRB2013}. Experimental studies have been done to understand mechanisms of epitaxial graphene growth on SiC(000$\bar{1}$)~\cite{BorysiukPRB2012}, SiC(0001)~\cite{SunPRB2011}, non-polar SiC surfaces~\cite{OstlerPRB2013} and to elucidate role of carbon diffusion~\cite{OhtaPRB2010} and silicon sublimation~\cite{SunPRB2011}. It has been shown that graphene with reduced pit density can be grown on nominally flat SiC substrates~\cite{SunPRB2011} and graphene quality can be further improved when grown in high argon pressure~\cite{BolenJEM2010}. On the other hand, the carrier mobility has been observed to decrease with increasing argon pressure when time and temeperature are kept constant~\cite{BolenJEM2010}. Other strategies to improve graphene quality involve thermal decomposition of deposited polymer adsorbate which acts as a carbon source~\cite{Kruskopf2DMaterials2016}. The SiC step bunching, as another issue reducing graphene mobility on SiC, has been solved by amorphous carbon step pinning~\cite{PalmerAPL2014}. The thermodynamics of stable phases that governs the onset of graphene formation~\cite{NemecPRB2015}, oxidation~\cite{HassDeHeer2008} and other chemical reactions~\cite{LuWeijieJPC2012} has been discussed, too. 
%%%RGA%%%%%%%%%%%%%%%%%%%%%%%%%%%%%%%%%%%%%%%%%%%%%%%%%%%%%%%%%%%%%%%%%%%%%%%%%%%%%%%%%%%%%%%%%%%%%%%%%%%%%%%%%%%%%%%
However, there is a little or no experimental evidence of graphene growth conditions and composition of a residual gas inside a graphene furnace~\cite{WeijieJoPD2010}. These studies are restricted to gas phase dynamics during Chemical Vapour Deposition (CVD) graphene growth~\cite{KwakPCCP2013} and to carbon nanotubes growth~\cite{PattinsonACSNano2012}.
%%%What we did new%%%%

Here, we determine experimentally the residual gas content in a furnace for epitaxial graphene growth on SiC. The measurements are performed in-situ by residual gas analyzer during whole graphene growth. We show that the initial graphene growth is accompanied by chemical reaction of residual water and carbon. The results of Norimatsu~\cite{NorimatsuPRB2011} indicate that control of the number of graphene layers require precise control of the first stage of SiC decomposition. We discuss that hydrogen, one of the products of the reaction of residual water and carbon, might be one reason for SiC step bunching during initial graphene growth stage. We compare the chemistry of graphene growth in high vacuum and in low argon pressure. The comparison of both growth techniques demonstrates that argon reduces the rate of water and carbon reaction. Therefore amount of hydrogen is smaller and graphene quality is higher. We discuss also the role of argon in terms of growth temperature, time and gas purity.
The graphene quality is analyzed by Raman spectroscopy mapping and by Hall effect measurements. We compare growth in vacuum and high pressure argon in terms of charge density, strain, graphene grain size and their spatial variation. 

\section{Samples and experimental setup}
The wafers of 6H-SiC were purchased from II-VI Inc. and diced into 3.5$\times$3.5~mm$^2$ rectangles. The samples are (500$\pm$25)~$\mu$m thick, semi-insulating high resistivity ($10^{10}~\Omega$cm) vanadium doped SiC. Graphene is grown on the silicon face SiC(0001) with a wafer miss-cut $\pm0.6^{\circ}$. The wafers were chemically-mechanically polished and the surface roughness is about 0.5~nm. Diced samples were rinsed and sonicated in both aceton and iso-propylalcohol for 10 minutes. The semi-insulating samples were heated in a graphite crucible by radio-frequency (RF) induction heating at 250-270~kHz. The graphite used for fabrication of the crucible is iso-statically pressed (ISP) graphite and glassy (vitreous) graphite. The ISP graphite has density 1850~kg/m$^3$ and electrical resistivity 11$\times10^{-4}~\Omega$cm. The glassy graphite, purchased from Final Advanced Materials, has been used as a reference to study effect of crucible pre-baking. The high epitaxial graphene quality has been reached by Confinement Controlled Sublimation (CCS) technique~\cite{deHeerPNAS2011}. The processing environment is either high vacuum $5\times10^{-6}$~mbar or high purity argon (6N purity gas vessel). The argon is further purified to the level of 9N purity by gas purifiers to remove residual oxygen and water. The stainless steel gas tubes have been outgassed at temperature about 200$^{\circ}$C for 2 hours prior to graphene growth to avoid contamination by residual water and impurities from the stainless steel surface. The temperature is monitored by either type C thermocouple or by two-color pyrometer working at 0.95 and 1.05~$\mu$m wavelengths. The pyrometer allows to measure temperature with spatial resolution of about 0.2~mm. The thermocouple type C is electrically isolated by alumina shielding. The graphite crucible is placed inside a hot wall reactor made of 4~mm thick and 250~mm long quartz tube. The quartz tube is opened from both sides to allow effective gas circulation. The growth environment has been studied by residual gas analyzer (Prisma Plus QMG 220). The CF-40 flanges are used for all connections except of the adapter between the furnace and the turbo-molecular pump and quartz tube feedthrough. The quartz feedthrough is made of two o-rings on both sides to allow leak-tight connection up to high vacuum $1.5\times10^{-6}$~mbar and yet allowing easy access to the reaction chamber.
\begin{figure}[t]
\centering
\includegraphics[width=6cm]{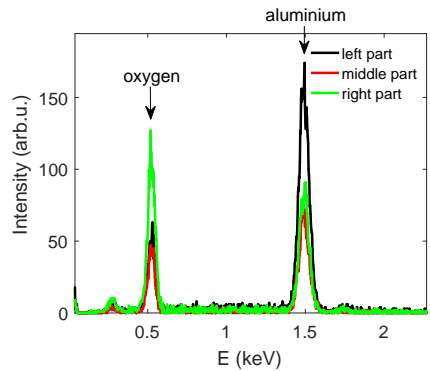}
\caption{Energy-dispersive X-ray analysis of quartz tube coated during the thermocouple monitored growth process. The aluminium-oxygen mixture with varying content of aluminium has been observed along the quartz tube.}
\label{FigEDS}
\end{figure} 
\section{Temperature measurements}
The standard furnace design for epitaxial graphene growth uses either thermocouple or pyrometer to monitor growth temperature. We have used both methods and compared them. The thermocouple is a cheaper method and it measures temperature inside the graphite crucible. It requires electrical isolation (made of alumina for high temperature purposes up to 2000$^{\circ}$C) complicating furnace design and introducing other elements beside carbon and silicon (SiC sample and C based crucible). We have found that the electrical shield made of alumina (polycrystalic Al$_2$O$_3$) is sputtered and deposited on a relatively colder quartz tube during cool down process. Energy-dispersive X-ray spectroscopy (EDS) has been performed to determine chemical composition of the sputtered material, as shown in Fig.~\ref{FigEDS} for three different positions along the inner part of quartz tube. We have found presence of aluminium and Al$_2$O$_3$ mixture. Their relative ratio changes along the quartz tube. As we have used thermocouple type C, which is composed of rhenium and tungsten, we have looked carefully for the traces of these metals as well. The negative result of our search has led us to conclusion that the sputtered alumina from the thermocouple electrical shield is a source of furnace contamination, very often seen in high temperature graphene furnaces. We expect alumina traces can also contaminate SiC wafer and graphene. Hence, in order to keep high graphene quality, we propose to use only remote temperature sensing by pyrometer. We also note it is essential to use two-color ratio pyrometer for the reason of possible contamination of furnace quartz windows. The contamination leads to decreased light intensity and to the lower effective temperature determined by one-color pyrometer. 
\begin{figure}[t]
\centering
\includegraphics[width=7cm]{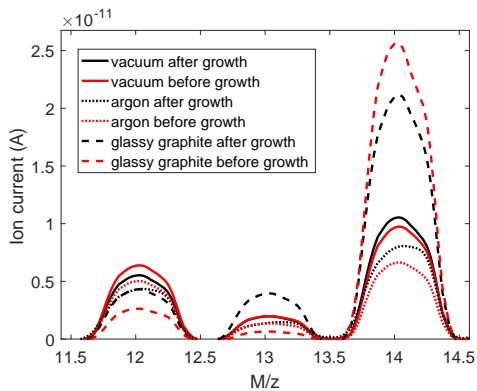}
\caption{Mass spectra between M/z=12 to 14. The residual gas spectra measured in the furnace before (red solid, dashed and dotted lines) and after (black solid, dashed and dotted lines) graphene growth at 1550$^\circ$C for 30~minutes. The pre-baking at 800$^\circ$C for 10~minutes and at 1000$^\circ $C for 10~minutes has been performed prior to graphene growth. The growth atmosphere has been studied for the case of high vacuum at $5\times10^{-6}$~mbar using graphite crucible (solid lines), low pressure 9N purity argon at $5\times10^{-6}$~mbar (dotted lines) and at high vacuum $5\times10^{-5}-10^{-4}$~mbar using unbaked glassy graphite crucible (dashed lines). The base pressure was $1.5\times10^{-6}$~mbar prior to RF heating in all cases.} 
\label{FigRGA}
\end{figure} 
\section{Graphite crucible}
We study here the effect of graphite crucible gas adsorption. The commonly used ISP graphite and glassy graphite crucibles are compared. The major issue of ISP graphite is its high porosity and permeability to gases. The high permeability leads to undesired contamination by air when samples are exchanged or graphite crucible is stored at ambient conditions for longer time (order of hours and longer). A glassy graphite is impermeable to gasses hence it is more convenient to keep the furnace environment clean of air residual gasses. The residual gas stored in ISP graphite is removed within the first two pre-bake graphene growth stages. The pressure usually grows from $1.5\times10^{-6}$~mbar to $\times10^{-3}$~mbar or more if crucible is not well baked. The well baked crucible exhibits small rise of the residual gas pressure by less than one order of magnitude, typically $1.5\times10^{-6}$~mbar to $\times10^{-5}$~mbar. Contrary to ISP graphite, glassy graphite is impermeable to gasses and the pre-baking removes surface absorbants only. The glassy carbon crucible is less susceptible to long exposure to air, however, its pre-baking at temperature much higher than the growth temperature is also necessary, as is shown in Fig.~\ref{FigRGA}. The well-baked graphite crucible exhibits comparable strength of peaks at M/z=12 (carbon) and 14 (nitrogen), which is also signature of leak-tight vacuum chamber. The unbaked glassy graphite exhibits more than 2$\times$ stronger nitrogen peak which is supposed to come from glassy graphite surface. We use peak at M/z=14 instead of M/z=28 because the latter one overlaps with carbon monoxide. 

\begin{figure}[t]
\centering
\includegraphics[width=9cm]{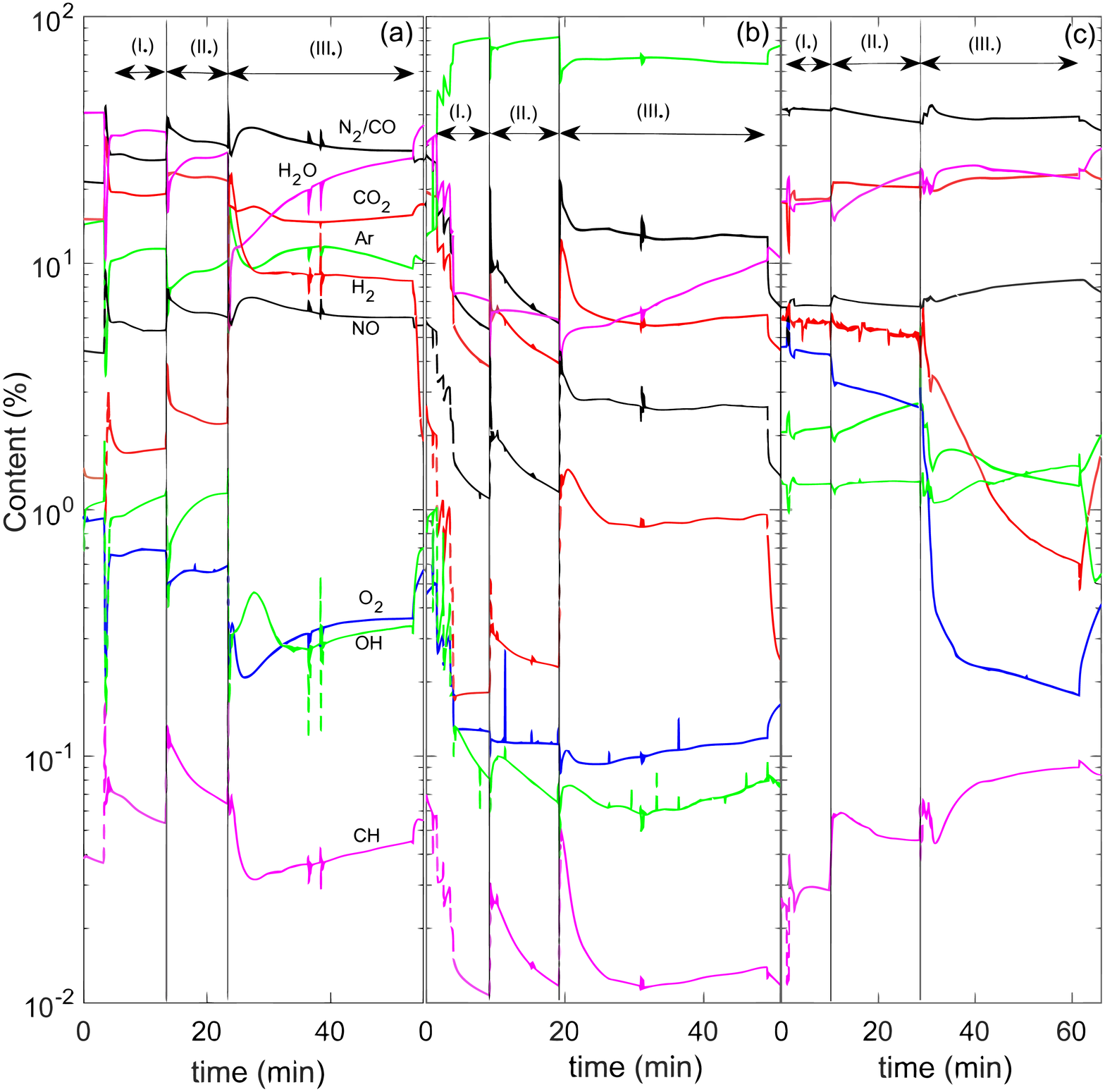}
\caption{Residual gas composition during graphene growth in high vacuum at $5\times10^{-6}$~mbar using graphite crucible (a), low pressure 9N purity argon at $5\times10^{-6}$~mbar (b) and at high vacuum $5\times10^{-5}-10^{-4}$~mbar using unbaked glassy graphite crucible (c). The three growth stages are marked as following: (I.) prebake at 800$^\circ$C for 10~minutes, (II.) prebake at 1000$^\circ$C for 10~minutes, (III.) graphene growth at 1550$^\circ$C for 30~minutes. The temporal evolution is shown for major growth environment components; nitrogen/carbon monoxide (black solid line), carbon dioxide (red solid line), argon (green solid line), oxygen (blue solid line), water (magenta solid line), nitrogen monoxide (black dashed line), hydrogen (red dashed line), -OH fragment (green dashed line), -CH fragment (magenta dashed line).}
\label{FigMCD}
\end{figure}
\begin{figure}[t]
\centering
\includegraphics[width=8cm]{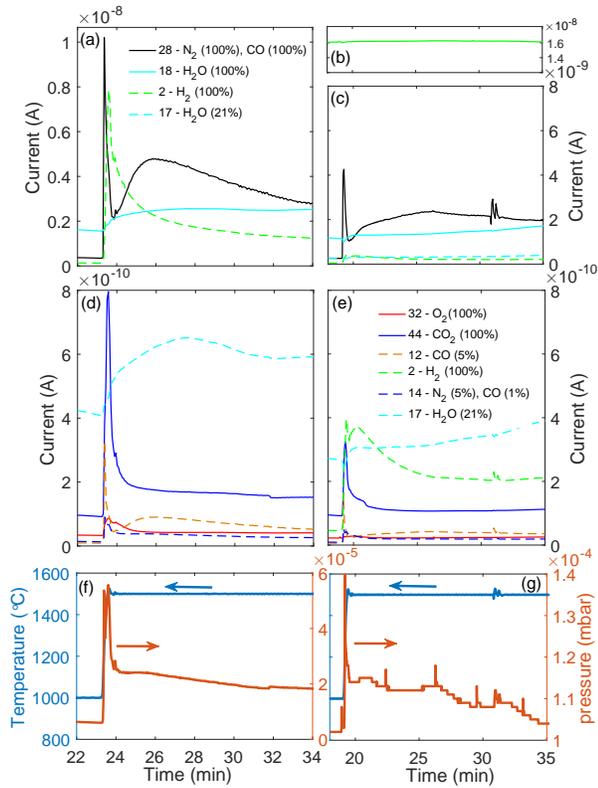}
\caption{The ion current for M/z ratios 2, 12, 14, 17, 18, 28, 32 and 44 in first 10~minutes of graphene growth at 1500$^{\circ}$C in high vacuum at 1-2~$\times10^{-5}$~mbar (a,d) and total pressure and temperature in shown in part (f). The M/z ion current is compared with low argon pressure growth at 1.1$\times10^{-4}$~mbar (9N purity) in part (b,c,e) and corrensponding pressure and temperature time evolution is shown in part (g).} 
\label{FigRGAdetail}
\end{figure}
\begin{figure}[t]
\centering
\includegraphics[width=7cm]{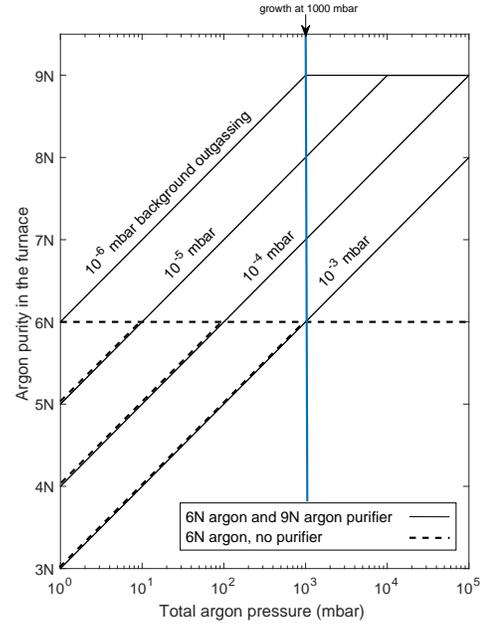}
\caption{A diagram of the growth environment purity as a function of total argon pressure in the furnace. A case of 6N pure argon is plotted by dashed lines and the case of 9N (purified from 6N) is plotted by solid lines. The final gas purity is calculated for four different background pressures, as is shown by labels $10^{-3}$~mbar, $10^{-4}$~mbar, $10^{-5}$~mbar and $10^{-6}$~mbar in diagram. }
\label{FigPurity}
\end{figure}

\section{Growth atmosphere}
The general growth process consist of three stages. The pre-baking stage (I.), depicted in Fig.~\ref{FigMCD}, is used for crucible outgassing at 800$^{\circ}$C for 10~minutes. The second pre-baking stage (II.) at 1000-1100$^{\circ}$C for 10~minutes is used to decompose native oxide layer on SiC wafer. The graphene growth stage (III.) is performed at temperatures 1400-2000$^{\circ}$C at times in the range of minutes to hours. Fig.~\ref{FigMCD} depicts relative content of growth ambient in all three growth stages for the case of growth in high vacuum (a) under continuous vacuum pumping by a turbo-molecular pump, low pressure argon (b) under continuous pumping and small argon leak and growth in high vacuum using glassy graphite crucible (c). The graphene growth stage (III.) has been kept at 30~minutes to show long-time evolution of the growth ambient. A change of relative content is caused by different pumping rate (e.g. slow hydrogen pumping), outgassing from vacuum chamber walls (mainly water) and chemical reactions. We have identified that oxidizing reactions appear at every temperature rise (from step (I.) to (II.) and from step (II.) to (III.)) resulting in reduced both relative and absolute amount of oxygen. These reactions are however negligible for very small content of oxygen in the chamber. The major identified reaction is a reaction of two major components in the furnace, carbon and water. We discuss this process in detail here. The epitaxial graphene can be grown fast (2-5 minutes) at higher temperatures or slowly (20-90 minutes) at lower temperatures. We have found that 1430$^{\circ}$C for 20~minutes is equivalent to 1500$^{\circ}$C for 5~minutes by amount of grown graphene. It is however not equivalent in growth conditions concerning the growth environment. We show difference between the two regimes. The main issue is a content of water and carbon monoxide, as shown in Fig.~\ref{FigRGAdetail} at the beginning of graphene growth. A water content is about 10-15\% for times shorter than 5 minutes, but it grows then steadily and it completely dominates the growth atmosphere at times longer than 30 minutes. A detail of growth ambient in high vacuum in Fig.~\ref{FigRGAdetail}~(a,d,f) shows sharp pressure rise accompanied by sharp rise of ion current at M/z=28 by 8$\times10^{-9}$~A. Similarly, the ion current increases at masses M/z=44 by 0.6$\times10^{-9}$~A and M/z=12 by 0.25$\times10^{-9}$~A. A significant rise of hydrogen is also present by amount of 6$\times10^{-9}$~A. The water ion current is unchanged within first 60~s of growth stage (III.). The step rise of ion current is expected instead due to the step rise of temperature and consequent step rise of mean velocity of ions. The lack of such ion step rise of water can be attributed to the reduction of total amount of water in the chamber. The rise of M/z=28 by 8$\times10^{-9}$~A is attributed to CO. The rise of CO is confirmed by rise of ion current at M/z=12 by 0.25$\times10^{-9}$~A, which is 3\% (expected CO contribution is 5\%) of the major ion current rise at M/z=28. The minor ion current peak at M/z=14 by 0.05$\times10^{-9}$~A (0.6\% of M/z=28, expected CO contribution is 1\%) is also attributed to CO. We conclude that there is no absolute change of nitrogen (M/z=28, 14) and all content changes can be explained by increasing amount of CO and similar amount of hydrogen. The shift of hydrogen peak towards longer times is caused by slower pumping of hydrogen than pumping of CO, water and other heavier ions. The ion current rise at mass M/z=44 by 0.6$\times10^{-9}$~A is due to the carbon dioxide (CO$_2$). Therefore the dynamics of growth ambient is dominated by chemical reaction
\begin{equation}
C(s)+H_2O(g)\rightleftharpoons CO(g)+H_2(g)+\mathrm{7.5\%}(CO_2(g)+CH_4).
\label{Eqreaction}
\end{equation}
The major chemical reaction given by equation~(\ref{Eqreaction}) is observed mainly on the short time scale below 30~s after each temperature step rise. The time scale between 30-100~s is dominated by pumping the products of Eq.~(\ref{Eqreaction}) out of the furnace. The long time scale above 2~minutes is dominated by both pumping of the residual gas and water outgassing from the stainless steel furnace walls. This behavior has been observed both in high vacuum and low pressure argon and also in the case of glassy graphite crucible in high vacuum, as is shown in Fig.~\ref{FigMCD}. The content changes only slowly for longer times ($>$60~s), however, water becomes slowly dominant gas. The water source is at the surface of stainless steel furnace walls. The main issue is that it can promote reaction with carbon (\ref{Eqreaction}) even further, hence creating defects in a graphene layer. Therefore it is more preferential to grow graphene at higher temperatures for shorter times.

As the residual gas analysis has shown preferable shorter growth times and higher temperatures, we grow single layer graphene between 1500-1550$^{\circ}$C for 5~minutes in high vacuum $5\times10^{-6}$-$2\times10^{-5}$~mbar. The 1500$^{\circ}$C/5~minutes growth seems to result in a patchy graphene with about 20-40\% of buffer layer. The single layer graphene is proven by Raman spectroscopy and discussed later. There are already signatures of bilayers at higher temperature 1550$^{\circ}$C/5~minutes. 

Growth in 800-1050~mbar of argon is approximatelly 30-50$^{\circ}$C shifted towards higher temperatures with respect to growth in high vacuum. Buffer grows at about 1410-1430$^{\circ}$C in high vacuum and at about 1450$^{\circ}$C at 800-1050~mbar argon pressure. The growth has been always done in the same graphite crucible with a hole 1~mm in diameter and 10~mm long. Argon acting on graphene growth is twofold. First, it is equivalent to CCG method, where the lowered silicon sublimation is due to the small volume in nearly closed graphite crucible. Second, argon pressure acts as an inhibitor for water outgassing, as has been discussed in previous paragraphs as the major source of residual gas and potential cause of defects formation in epitaxial graphene. We note, that high purity argon is necessary in order to obtain overall purity of growth conditions comparable to those in high vacuum at $1\times10^{-5}$~mbar. An equivalent purity of argon gas for graphene growth at 1000~mbar is 8N (=1000~mbar/10$^{-5}$~mbar). A lower argon purity results in higher relative content of residual gas, which leads to higher total amount of mainly water at 1000~mbar argon pressure growth (the total amount of water is discussed with respect to the total amount of water at high vacuum (10$^{-5}$~mbar) growth). Beside the gas purity, also the background outgassing in the furnace has to be taken into account. A typical water backround pressure in unbaked furnace is 10$^{-6}$~mbar-10$^{-5}$~mbar. If the background water pressure is higher, the overal argon purity in the furnace will be lower even with gas purifiers, as is described in Fig.~\ref{FigPurity}. Fig.~\ref{FigPurity} demonstrates overall gas purity as a function of total argon pressure for two degrees of argon purity. The solid (dashed) lines depict total argon purity as a function of total argon pressure for inlet of 9N (6N) pure argon. The total gas purity cannot be higher than the inlet gas purity, therefore all curves saturate at the value of inlet gas purity 9N (solid lines) and 6N (dashed lines). Fig.~\ref{FigPurity} also shows influence of four different background impurity pressures. The background impurity pressure is detrimental at low total argon pressures, therefore it is expected that higher quality graphene will be grown at higher argon pressures, as has been evidenced in literature~\cite{BolenJEM2010}. A further reduction of water outgassing could be done by growth in argon flow. The graphene quality should be independent on the flow rate as long as the flow rate is much higher than outgassing rate. Another important role of argon is reduction of water and carbon reaction by factor of three, as can be seen in Fig.~\ref{FigRGAdetail}~(c,e,g), as compared to growth in high vacuum, Fig.~\ref{FigRGAdetail}~(a,d,f). It can be assumed that the reaction rate will be reduced even further at higher argon pressure.

\begin{figure}[t]
\centering
\includegraphics[width=7cm]{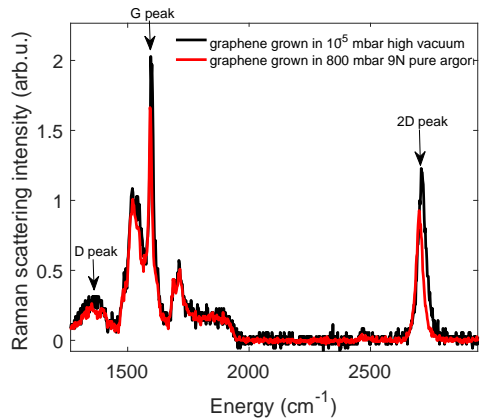}
\caption{Raman spectra of graphene grown at 10$^{-5}$~mbar (black curve) and in 9N pure argon at 800~mbar (red curve). The positions of D, G and 2D peaks are marked by labeled arrows.}
\label{FigRamanSpectra}
\end{figure}
\begin{figure}[h]
\centering
\includegraphics[width=9cm]{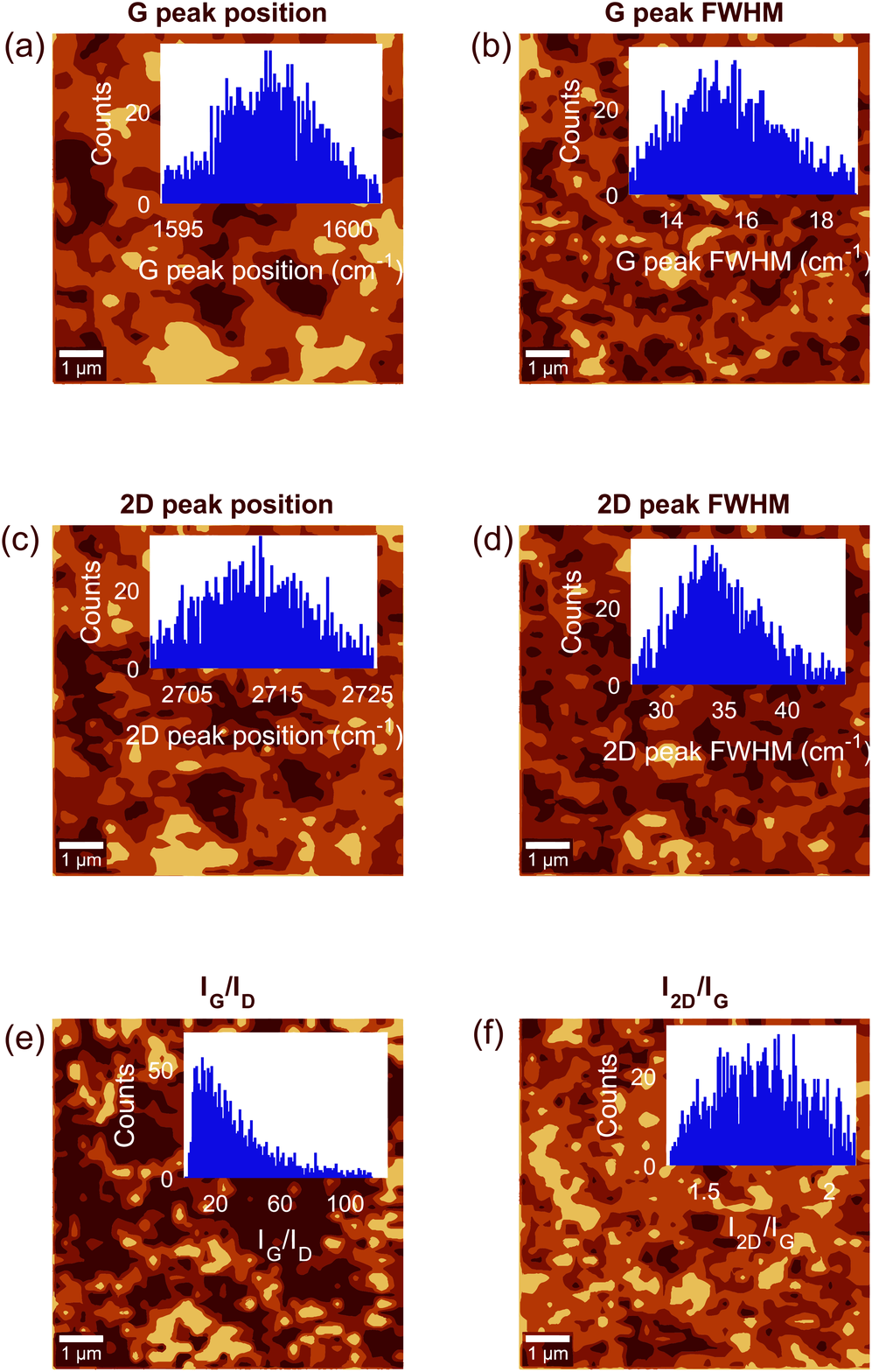}
\caption{Maps of Raman spectra parameters shown for a characteristic graphene sample grown in high vacuum at $5\times10^{-6}-10^{-5}$~mbar at 1550$^\circ$C for 5~minutes. The analyzed Raman spetra parameters are G peak position (a), G peak FWHM (b), 2D peak position (c), 2D peak FWHM (d), ratio of integrated G peak and D peak intensities (e) and ratio of integrated 2D peak and G peak intensities. The scale bar $1~\mu$m is shown in the left bottom corner in each map. The statistical analysis is depicted in a form of histogram for each analyzed parameter including units and the range of values used in color scales. }
\label{FigRamanVac}
\end{figure}
\begin{figure}[h]
\centering
\includegraphics[width=9cm]{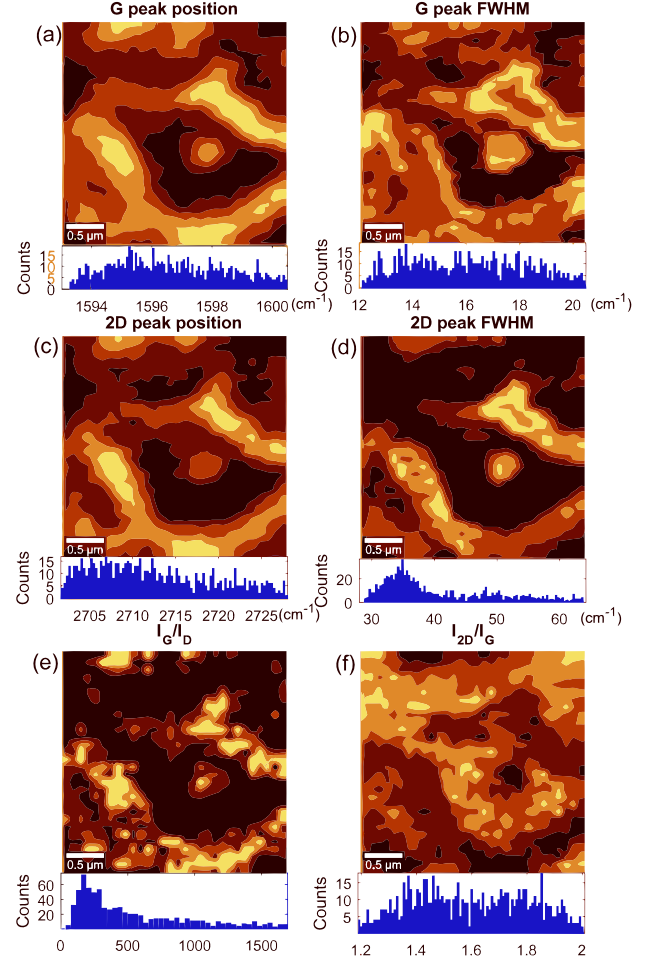}
\caption{Maps of Raman spectra parameters shown for a characteristic graphene sample grown in ultra-pure argon (9N purity) at 800~mbar at 1600$^\circ$C for 5~minutes. The analyzed Raman spetra parameters are G peak position (a), G peak FWHM (b), 2D peak position (c), 2D peak FWHM (d), ratio of integrated G peak and D peak intensities (e) and ratio of integrated 2D peak and G peak intensities. The scale bar $1~\mu$m is shown in the left bottom corner in each map. The statistical analysis is depicted in a form of histogram for each analyzed parameter including units and the range of values used in color scales.}
\label{FigRamanArgon}
\end{figure}
\begin{figure}[h]
\centering
\includegraphics[width=9cm]{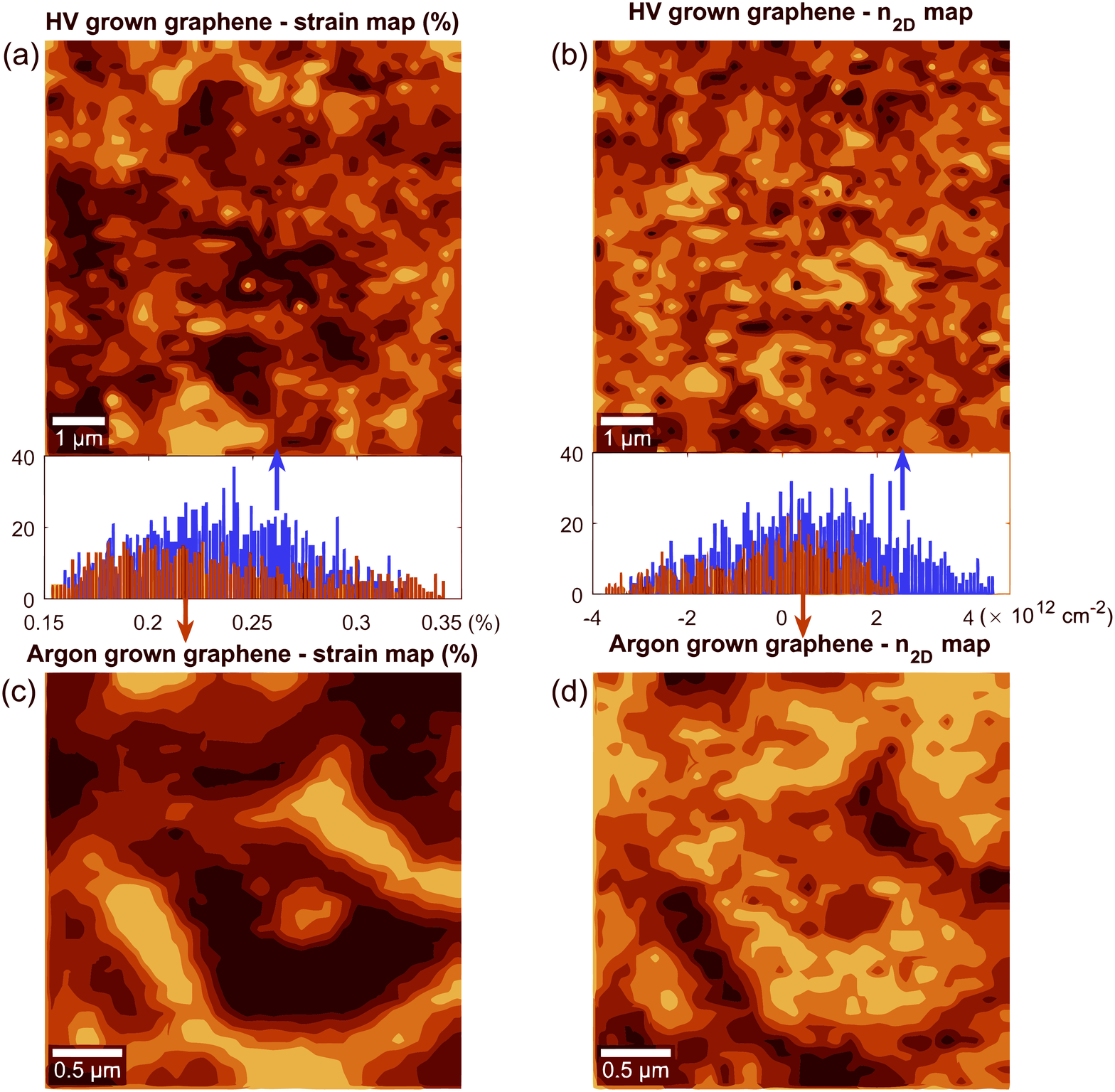}
\caption{The spatial distribution of mechanical strain (a,c) and charge density (b,d) in epitaxial graphene grown in high vacuum $5\times10^{-6}-10^{-5}$~mbar (a,b) and 800~mbar 9N-argon atmosphere (c,d). The statistical distribution of strain and charge density in both types of graphene samples is shown in corresponding histograms for high vacuum grown graphene (blue histograms) and high pressure argon grown graphene (red histograms). The scale bars are shown in the left bottom corner of each map; 1~$\mu$m for high vacuum grown graphene and 0.5~$\mu$m for high pressure argon grown graphene.}
\label{FigStrainChargeMap}
\end{figure}
\section{Raman spectroscopy}
We compare here two epitaxial graphene samples grown in high vacuum (HV-graphene) at 10$^{-5}$~mbar and in 9N pure argon at 800~mbar (Ar-graphene). First, Raman spectroscopy has been performed in order to determine strain, charge density distribution and graphene grain size. We have used confocal microscope with 100$\times$ magnification microscope objective and numerical aperture $NA=0.9$. The excitation laser wavelength was 532~nm and the laser power at the sample was 5~mW. The characteristic Raman spectra of HV-graphene and Ar-graphene are shown in Fig.~\ref{FigRamanSpectra}. Nearly absent D-peak shows on better graphene quality grown in ultra pure argon atmosphere. We have analyzed the positions, Full Width at Half Maximum (FWHM) and integrated intensities $I_D$, $I_G$ and $I_{2D}$ of D, G and 2D peak, respectively. We have constructed spatial maps of G and 2D peak positions, G and 2D peak FWHM and spatial maps of $I_G/I_D$ and $I_{2D}/I_G$, as shown in Fig.~\ref{FigRamanVac} and Fig.~\ref{FigRamanArgon} for HV-graphene and Ar-graphene, respectively. The positions of G and 2D peak has been calculated as Center of Mass (COM) of each peak after subtructing SiC background signal. The larger grain size of Ar-graphene can be seen from the fine structure of all spatial maps. We have used these spatial maps further to determine mechanical strain $\epsilon$
\begin{equation}
\epsilon=\frac{\omega_{2D}-\omega_{2D,0}}{2\omega_{2D,0}\gamma_{2D}},
\end{equation}
given by measured position of 2D peak $\omega_{2D}$, position of 2D peak at zero strain $\omega_{2D,0}=2677$~cm$^{-1}$ and Gr{\"u}neisen parameter for 2D peak $\gamma_{2D}=2.8$~\cite{SchmidtPRB2011}. The effects of charge density on the position of 2D peak are negligible~\cite{StampferAPL2007}. The charge density $n_{2D}$ is determined from G peak position $\omega_G$ and strain $\epsilon$,
\begin{equation}
n_{2D}=\alpha\left[(\omega_G-2\gamma_G\omega_{G,0}\epsilon)-\omega_{G,0}\right],
\end{equation}
where $\gamma_G$ is Gr{\"u}neisen parameter for G peak, $\omega_{G,0}=1582$~cm$^{-1}$ is G peak position for unstrained ($\epsilon=0$) and neutral (n$_{2D}$=0) graphene. Parameter $\alpha$=10$^{12}$~cm$^{-2}$/1~cm$^{-1}$ is approximately a proportionality factor between G peak position and graphene charge density, as determined theoretically~\cite{LazzeriPRL2006} and confirmed experimentally~\cite{DasNatureNanotech2008,StampferAPL2007}. Graphene grain size $L_a$ can be determined from the ratio $I_G/I_D$~\cite{Dresselhaus2010,CancadoNanoLetters2011,CancadoAPL2006,FerrariNatureNanotech2013} of integrated G peak intensity ($I_G$) and D peak intensity ($I_D$),
\begin{equation}
L_a(\mathrm{nm})=2.4\times10^{-10}\lambda^4(\mathrm{nm})\frac{I_G}{I_D},
\end{equation}
giving $L_a=19~$nm$\times\frac{I_G}{I_D}$ for excitation wavelength 532~nm.

Number of graphene layers has been determined by several authors~\cite{HwangNanotech2013} from the ratio of integreated 2D peak $I_{2D}$ and integrated G peak intensity $I_G$. This ratio $\mathcal{N}=I_{2D}/I_G=2$ for single layer graphene, $\mathcal{N}=1$ for bilayer and it is $\mathcal{N}<1$ for more than two graphene layers. We note this relation is the most discutable and it is probably valid only in the limit of few graphene layers. It cannot be used e.g. for number of graphene layer determination of multilayer graphene on C-face of SiC. Our graphene samples have been grown about 50$^\circ$C above the lowest graphitization temperature at 5~minutes and graphene has been always grown on Si-face of SiC, therefore we are in the limit of 1 to 2 layers and this method can be used to determine number of graphene layers. Fig.~\ref{FigRamanVac}~(f) and Fig.~\ref{FigRamanArgon}~(f) show spatial map of the ratio $I_{2D}/I_G$ including statistical analysis  in a form of histogram. Both samples consist of mostly single layer. A 2D-peak FWHM is (35$\pm$3)~cm$^{-1}$ in both samples. The narrow 2D peak is also a fingreprint of a single layer graphene~\cite{LeeSmetNanoLetters2008} and FWHM values about 30~cm$^{-1}$ has been correlated with charge mobility of about 10~000~cm$^{2}$V$^{-1}$s$^{-1}$ at Dirac point~\cite{RobinsonNanoLett2009}. The grain size in HV-graphene is $L_{a,HV}=(400\pm200)$~nm and it reaches $L_{a,Ar}=(1.5\pm0.5)$~$\mu$m in Ar-graphene. Strain is $\epsilon_{HV}=(0.24\pm0.05)$~\% and $\epsilon_{Ar}=(0.21\pm0.04)$~\% in HV and Ar grown graphene, respectively. The expected strain $\epsilon_{theory}=\frac{13a_{0,G}-6\sqrt{3}a_{0,SiC}}{13a_{0,G}}=0.19\%$ is determined from graphene ($a_{0,G}=2.462~\mathrm{\AA}$)~\cite{PozzoPRL2011} and SiC ($a_{0,SiC}=3.073~\mathrm{\AA}$) lattice constants at 300~K assuming $6\sqrt{3}\times6\sqrt{3}R30^{\circ}$ SiC-graphene super-cell~\cite{EmtsevPRB2008} containing 13 graphene unit cells and mutual rotation of graphene and SiC lattices by 30$^{\circ}$. Both samples show low doping level in the range $0.2-4\times10^{12}$~cm$^{-2}$ and the statistically significant amount of low density areas show on electron-hole puddles~\cite{AzarPRB2011}. A small doping is also evidenced by G peak FWHM in the range of 14-18~cm$^{-1}$~\cite{YanPRL2007}.
\begin{figure}[h]
\centering
\includegraphics[width=5cm]{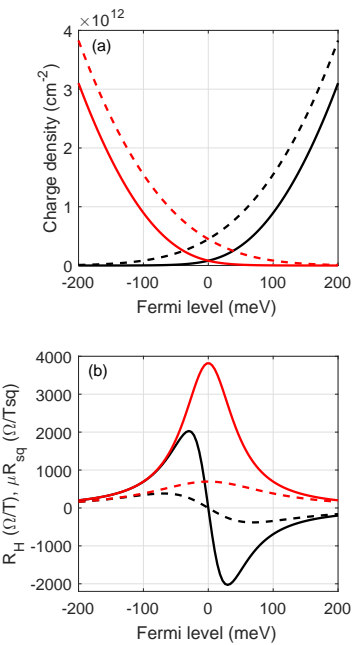}
\caption{Charge transport properties of pristine graphene (solid lines) and for graphene in fluctuating electrostatic potential (dashed lines) as a function of mean Fermi level $\bar{E}_F$. The charge density of electrons (black lines) and holes (red lines) is shown in part (a). The Hall coefficient $R_H$ (black lines) and $\mu R_{\Box}$ product (red lines) are shown in part (b). The standard deviation of the electrostatic potential is taken to be $\delta V=100$~mV (all dashed curves).}
\label{FigHallCoef}
\end{figure}

\section{Hall measurements}
We have measured Hall effect to determine charge density and carrier mobility. The room temperature measurements have been performed in magnetic field up to $\pm$0.35~T in van der Pauw configuration. A direct current (DC) source has been used ($I=1~\mu$A) and all combinations of current directions and voltages have been measured to determine specific resistance and Hall coefficient. Typical transport coefficients are shown in Tab.~\ref{TabTransport}. We have observed twice lower specific resistance and 1.8$\times$ higher apparent carrier mobility of argon grown graphene.
\begin{table}[t!]
\begin{tabular}{l|c|c|c|c}
       & $R_{\Box}$   & $R_{H}$  & $1/eR_H$       & $R_H/R_{\Box}$  \\
       &  (k$\Omega$/$\Box$) &  ($\Omega$/T) & (10$^{12}$)~cm$^{-2}$ & (cm$^{2}$V$^{-1}$s$^{-1}$)\\
\hline
HV-graphene    &  2400$\pm$100  &  -96$\pm$4      &  -6.5$\pm$0.5   &   390$\pm$20   \\
Ar-graphene    &  1150$\pm$50  &  -78$\pm$10      &  -8.0$\pm$1.0   &   690$\pm$80   \\
\end{tabular}
\caption{Transport properties of epitaxial graphene grown in high vacuum $5\times10^{-6}-10^{-5}$~mbar (HV-graphene) and 800~mbar 9N-argon atmosphere (Ar-graphene). The typical values are presented for specific two-dimensional resistivity (resistance per square) $R_{\Box}$, Hall coefficient $R_H$, apparent carrier density $1/eR_H$ and apparent carrier mobility $R_H/R_{\Box}$.}
\label{TabTransport}
\end{table}
We note that all samples show small negative Hall coefficient $R_{H}$ in the range from -50 to -100~$\Omega$/T, indicating n-type doping on the order of upper 10$^{12}$~cm$^{-2}$. Such charge densities are consistent with data in Ref.~\cite{TedescoGaskill2009}. However, it has been realized later~\cite{CurtinTedescoGaskillAPL2011} that instead of assuming single type conductivity, both electrons and holes should be considered in order to correctly interpret transport data exhibiting small Hall coefficient. This is so called model of electron-hole puddles~\cite{AzarPRB2011}. The electron-hole puddles have been observed previously by several authors~\cite{WiedmannPRB2011,CurtinTedescoGaskillAPL2011}. For this reason, we use notation of Ref.~\cite{CurtinTedescoGaskillAPL2011} and call the transport properties to be apparent charge density and apparent carrier mobility. The small Hall coefficient then leads to low carrier density. The low charge density is also consistent with our Raman scattering measurements, as shown in Fig.~\ref{FigStrainChargeMap}. The presence of electron-hole puddles leads to very high carrier mobilities above ~10~000~cm$^2$V$^{-1}$s$^{-1}$ for pristine graphene at 300~K. More realistic value can be estimated by considering electrostatic potential fluctuations. Electrostatic potential fluctuations smear out the Fermi level and increase an intrinsic charge density. We assume normal potential distribution with standard deviation $\delta V$. The charge density is calculated for given Fermi level $E_F$ by
\begin{equation}
n_{2D}=\frac{2}{\pi\hbar^2v_F^2}\int_0^\infty\frac{EdE}{e^{\frac{E-E_F}{k_BT}}+1},
\label{Eqn2D}
\end{equation}
where $v_F$ is Fermi velocity. The mean charge density $\bar{n}_{2D}$ caused by the potential fluctuations is calculated as a weighted sum
\begin{equation}
\bar{n}_{2D}(\bar{E}_F)=\int_0^\infty n_{2D}(E_F)\frac{1}{\sqrt{2\pi}\delta V}e^{-\frac{(E_F-\bar{E}_F)^2}{2\delta V^2}}dE_F.
\end{equation}
The mean hole charge density $\bar{p}_{2D}$ has been calculated by considering electron-hole symmetry. The mean electron and hole densities are plotted as a function of mean Fermi level $\bar{E}_F$ in Fig.~\ref{FigHallCoef}~(a) by solid lines for pristine graphene ($\delta V=0$~mV) and by dashed lines for graphene influenced by electrostatic potential fluctuations with standard deviation $\delta V=100$~mV. The total electrical current (two-dimensional, units (A/m)) is a sum of electron and hole currents.
\begin{equation}
j=j_e+j_h=e\bar{n}_{2D}\mu_eE+e\bar{p}_{2D}\mu_hE,
\end{equation}
where $\mu_e$ ($\mu_h$) is electron (hole) mobility. A specific resistivity, or resistance per square, is
\begin{equation}
R_{\Box}=\frac{1}{e(\bar{n}_{2D}\mu_e+\bar{p}_{2D}\mu_h)}.
\label{EqRsq}
\end{equation}
We plot a product of the specific resistance $R_{\Box}$ and mobility $\mu$ of pristine graphene and graphene influenced by electrostatic potential fluctuations ($\delta V=100$~mV) in Fig.~\ref{FigHallCoef}~(b). The $R_{\Box}$ is lowered for graphene influenced by electrostatic potential fluctuations because mean charge density (and intrinsic charge density) in graphene rises.
A Hall coefficient $R_H$ is given in two-dimensions and in the limit of $B\rightarrow0$~T by Eq.~(\ref{EqRH})
\begin{equation}
R_H=\frac{U_H}{I_xB_z}=\frac{p_{2D}\mu_h^2-n_{2D}\mu_e^2}{e(p_{2D}\mu_h+n_{2D}\mu_e)^2}.
\label{EqRH}
\end{equation}
The Hall coefficient is plotted in Fig.~\ref{FigHallCoef}~(b) for pristine graphene (solid black curve) and for the case of electrostatic potential fluctuations $\delta V=100$~mV by dashed black curve. The Hall coefficient is largely reduced for all mean Fermi level positions. The value  $\delta V=100$~mV is taken as an upper bound determined in other works~\cite{HuangPRB2015,RajputAPL2014,CurtinTedescoGaskillAPL2011,MartinNaturePhys2008}. A small measured Hall coefficient $R_H$ in the range of -50 to $-100~\Omega$/T suggests Fermi level in the range of 5-10~meV for $\delta V=100$~mV, which corresponds to about 5$\times$10$^{11}$~cm$^{-2}$ electron and 4$\times$10$^{11}$~cm$^{-2}$ hole density. The analysis of transport properties reduces now to determine $n_{2D}$, $p_{2D}$, $\mu_e$, $\mu_h$. Because we have determined $R_{\Box}$ and $R_H$ and estimated $\delta V$, we cannot determine any transport properties prior to further assumptions. The electron and hole charge densities are coupled by two common parameters, mean Fermi level $\bar{E}_F$ and electrostatic potential fluctuations $\delta V$. We assume equal electron and hole mobilities $\mu_e=\mu_h$, which reduces number of unknown parameters to three (being equal to three measured properties ($R_{\Box}$, $R_H$, $\delta V$). We use measured Hall coefficient $R_H$ to determine  mean Fermi level from Fig.~\ref{FigHallCoef}~(b). The mean Fermi level provides a value of $R_{\Box}\mu=690$~$\Omega/T$, which leads immediately to carrier mobility provided the specific resistance $R_{\Box}$ is known. We get carrier mobilities 2900 and 6000~cm$^{2}$V$^{-1}$s$^{-1}$ for vacuum and argon grown graphene, respectively. 

\section{Discussion}
The results of Raman scattering and Hall effect measurements lead us to conclusion that graphene grown in ultra-pure argon at 800~mbar is of higher quality than graphene grown in high vacuum. The growth in high vacuum leads to smaller graphene grain size below 1~$\mu$m even if the growth conditions are well optimized by high temperature bakeout of graphite crucible, gas inlet pipes to vent the furnace and the furnace itself. The key isssue in graphene growth is a presence of water and water outgassing from stainless steel surfaces. We have observed chemical reaction of water and solid state carbon in all stages of graphene growth. This reaction is important within first one minute of each graphene growth step. It becomes important again when graphene is grown for longer times above 20-30~minutes. This is caused by gradual furnace warm up and consequent water outgassing from previously colder furnace parts. Even though it would be thermodynamicaly more preferable to grow graphene slowly, providing high crystal quality, the water outgassing acts against long growth times. As we do not know the origin of carbon, we assume that the main source is a graphite crucible for a time scale below 60~s at the beginning of each growth step. However, as graphene grows, the carbon source can be also graphene itself. This reaction of water and carbon from partially formed graphene will be a cause of higher defect density and smaller grain size, as we have observed in Raman spectra analysis. A disadvantage of short growth time is that there is no temporally stable growth atmosphere content. Moreover, it differs slightly from one sample to another. The hydrogen produced in a reaction of carbon and water can also cause SiC etching or it can react with carbon, increasing defect density in graphene even more. Hydrogen is therefore undesired gas during graphene growth.
The growth in argon has been shown to lead to more stable growth atmosphere content with small rate of carbon and water reaction and also smaller amount of hydrogen. The high argon pressure reduces both water outgassing and also mean free path of water molecules. The shorter mean free path can be used to restrict water transport towards grown graphene when growth is performed in argon flow. We also point out that it is essential to use ultra-pure argon (8 or 9N) at the furnace inlet. A common 6N argon at 1000~mbar is in terms of total amount of residual gas equivalent to vacuum growth at $10^{-3}$~mbar (if background outgassing causes lower or equal pressure). Therefore, growth in 6N argon at 1000~mbar will lead to much worse graphene than the one grown in HV at $10^{-6}$-$10^{-5}$~mbar. The argon flow rate is not supposed to improve the overall amount of residual gas because this residual gas in homogeneously spread within the furnace volume. Contrary to that, the 9N argon at 1000~mbar is equivalent to HV at $10^{-6}$~mbar in terms of total amount of residual gas. These two growth conditions are initially equivalent. However, the growth conditions are more preferable for argon growth at the later stages of graphene growth for two reasons. First, water outgasses in HV and causes graphene degradation (pressure rises to 10$^{-5}$~mbar). Water ougassing is strongly reduced in high pressure argon, therefore the total amount of impurities remains nearly constant. Second, any possible residual gas in ultra pure argon can be removed by growth in argon flow. The argon flow should be only higher than the outgassing rate. 
Air permeability of graphite brings constrictions on a crucible storage. The nitrogen and water is absorbed into the crucible after longer exposure to air. It is therefore suggested to store ISP graphite in vacuum or inert gas. Glassy graphite is impermeable to air components, however, the surface absorbants still lead to higher desorption when glassy graphite crucible is heated to growth temperatures. 
Raman spectroscopy has proven to be a versatile tool to determine most of basic graphene properties as mechanical strain, charge density, graphene grain size, number of graphene layers or even to estimate electrostatic potential fluctuations and carrier mobility. We have measured graphene Raman spectra using several microscope objetives with different magnification and numerical apertures. We point out that although relative strength of SiC Raman spectrum and intensity of graphene spectrum (particularly intensity of G peak) have been used to determine number of graphene layers by some authors~\cite{ShivaramanJEM2009}, this method can be only calibrated (intensity ratio to the number of layers) to a given experimental set up and it is not transferable. The reason is mainly due to the different microscope objective numerical apertures (NA). The smaller the apperture is the longer the Rayleigh length. It causes that small NA objective probes signal deeper from bulk crystal, even if it is focused properly on a surface graphene layer. As a result, relative intensity of graphene to SiC bulk spectrum will be smaller than the ratio determined from Raman spectra measured with high-NA objective. A similar issue comes up in a grain size determination from the ratio of D to G peak intensity. If low-NA objective is used, graphene signal is weak realitively to SiC signal and the D peak can seem to be negligible with respect to SiC~\cite{LeeNanoLetters2008}. The small D peak is then mistakenly identified as a signature of high quality graphene.
Hall effect measurements have been used to quantify graphene electronic quality. The apparent carrier mobilites are almost twice higher in argon grown graphene. The analysis of transport measurements has been done also within a model of electron hole puddles. This model is in agreement with our Raman scattering analysis which has indicated very low charge density below 1.5~$\times$10$^{12}$~cm$^{-2}$. The carrier mobility is twice higher for argon grown graphene within the model of electron hole puddles, too. 

\section{Conclusion}
We have analyzed experimentally graphene growth environment and its influence on graphene quality. Residual gas analysis has shown important role of water reaction with carbon in high vacuum growth ambient. We have shown reduced role of the residual gas in 9N pure argon and we have compared high vacuum and argon grown graphene. The gas purity is a detrimental factor for high quality graphene for electronic applications. The mechanical strain, charge density, graphene grain size and number of graphene layers has been determined from Raman spectroscopy mapping. Hall measurements in van der Pauw configuration have provided information about charge density consistent with Raman spectroscopy. A carrier mobility has been modeled within electrostatic potential fluctuations model to provide realistic model of mean Fermi level dependence of Hall coefficient. The estimated lower bound of carrier mobilities in lithographically unprocessed epitaxial graphene provides a new route towards high-frequency electronic devices based on epitaxial graphene on SiC. 

\section{Acknowledgement}
Financial support from the Grant Agency of the Czech Republic under contract 16-15763Y is gratefully acknowledged. Raman spectroscopy has been measured within project VaVpI CZ.1.05/4.1.00/16.0340. We acknowledge technical support from K. Uhl\'{i}\v{r}ov\'{a} and fruitful discussion with P. Hl\'{i}dek and R. Grill.


\begin{thebibliography}{44}
\expandafter\ifx\csname natexlab\endcsname\relax\def\natexlab#1{#1}\fi
\expandafter\ifx\csname bibnamefont\endcsname\relax
  \def\bibnamefont#1{#1}\fi
\expandafter\ifx\csname bibfnamefont\endcsname\relax
  \def\bibfnamefont#1{#1}\fi
\expandafter\ifx\csname citenamefont\endcsname\relax
  \def\citenamefont#1{#1}\fi
\expandafter\ifx\csname url\endcsname\relax
  \def\url#1{\texttt{#1}}\fi
\expandafter\ifx\csname urlprefix\endcsname\relax\def\urlprefix{URL }\fi
\providecommand{\bibinfo}[2]{#2}
\providecommand{\eprint}[2][]{\url{#2}}

\bibitem[{\citenamefont{Berger et~al.}(2004)\citenamefont{Berger, Song, Li, Li,
  Ogbazghi, Feng, Dai, Marchenkov, Conrad, First et~al.}}]{Berger2004}
\bibinfo{author}{\bibfnamefont{C.}~\bibnamefont{Berger}},
  \bibinfo{author}{\bibfnamefont{Z.}~\bibnamefont{Song}},
  \bibinfo{author}{\bibfnamefont{T.}~\bibnamefont{Li}},
  \bibinfo{author}{\bibfnamefont{X.}~\bibnamefont{Li}},
  \bibinfo{author}{\bibfnamefont{A.}~\bibnamefont{Ogbazghi}},
  \bibinfo{author}{\bibfnamefont{R.}~\bibnamefont{Feng}},
  \bibinfo{author}{\bibfnamefont{Z.}~\bibnamefont{Dai}},
  \bibinfo{author}{\bibfnamefont{A.}~\bibnamefont{Marchenkov}},
  \bibinfo{author}{\bibfnamefont{E.}~\bibnamefont{Conrad}},
  \bibinfo{author}{\bibfnamefont{P.}~\bibnamefont{First}},
  \bibnamefont{et~al.}, \bibinfo{journal}{Jour. Phys. Chem. B}
  \textbf{\bibinfo{volume}{108}}, \bibinfo{pages}{19912}
  (\bibinfo{year}{2004}), ISSN \bibinfo{issn}{1520-6106}.

\bibitem[{\citenamefont{Kunc et~al.}(2014)\citenamefont{Kunc, Hu, Palmer, Guo,
  Hankinson, Gamal, Berger, and de~Heer}}]{KuncNanoLetters2014}
\bibinfo{author}{\bibfnamefont{J.}~\bibnamefont{Kunc}},
  \bibinfo{author}{\bibfnamefont{Y.}~\bibnamefont{Hu}},
  \bibinfo{author}{\bibfnamefont{J.}~\bibnamefont{Palmer}},
  \bibinfo{author}{\bibfnamefont{Z.}~\bibnamefont{Guo}},
  \bibinfo{author}{\bibfnamefont{J.}~\bibnamefont{Hankinson}},
  \bibinfo{author}{\bibfnamefont{S.~H.} \bibnamefont{Gamal}},
  \bibinfo{author}{\bibfnamefont{C.}~\bibnamefont{Berger}}, \bibnamefont{and}
  \bibinfo{author}{\bibfnamefont{W.~A.} \bibnamefont{de~Heer}},
  \bibinfo{journal}{Nano Lett.} \textbf{\bibinfo{volume}{14}},
  \bibinfo{pages}{5170} (\bibinfo{year}{2014}), ISSN \bibinfo{issn}{1530-6984}.

\bibitem[{\citenamefont{Dong et~al.}(2014)\citenamefont{Dong, Guo, Palmer, Hu,
  Ruan, Hankinson, Kunc, Bhattacharya, Berger, and
  de~Heer}}]{RuiJPDApplPhys2014}
\bibinfo{author}{\bibfnamefont{R.}~\bibnamefont{Dong}},
  \bibinfo{author}{\bibfnamefont{Z.}~\bibnamefont{Guo}},
  \bibinfo{author}{\bibfnamefont{J.}~\bibnamefont{Palmer}},
  \bibinfo{author}{\bibfnamefont{Y.}~\bibnamefont{Hu}},
  \bibinfo{author}{\bibfnamefont{M.}~\bibnamefont{Ruan}},
  \bibinfo{author}{\bibfnamefont{J.}~\bibnamefont{Hankinson}},
  \bibinfo{author}{\bibfnamefont{J.}~\bibnamefont{Kunc}},
  \bibinfo{author}{\bibfnamefont{S.~K.} \bibnamefont{Bhattacharya}},
  \bibinfo{author}{\bibfnamefont{C.}~\bibnamefont{Berger}}, \bibnamefont{and}
  \bibinfo{author}{\bibfnamefont{W.~A.} \bibnamefont{de~Heer}},
  \bibinfo{journal}{Jour. Phys. D-Appl. Phys.} \textbf{\bibinfo{volume}{47}}
  (\bibinfo{year}{2014}), ISSN \bibinfo{issn}{0022-3727}.

\bibitem[{\citenamefont{Yazdi et~al.}(2016)\citenamefont{Yazdi, Iakimov, and
  Yakimova}}]{YazdiCrystals2016}
\bibinfo{author}{\bibfnamefont{G.~R.} \bibnamefont{Yazdi}},
  \bibinfo{author}{\bibfnamefont{T.}~\bibnamefont{Iakimov}}, \bibnamefont{and}
  \bibinfo{author}{\bibfnamefont{R.}~\bibnamefont{Yakimova}},
  \bibinfo{journal}{Crystals} \textbf{\bibinfo{volume}{6}}
  (\bibinfo{year}{2016}), ISSN \bibinfo{issn}{2073-4352}.

\bibitem[{\citenamefont{Nemec et~al.}(2015)\citenamefont{Nemec, Lazarevic,
  Rinke, Scheffler, and Blum}}]{NemecPRB2015}
\bibinfo{author}{\bibfnamefont{L.}~\bibnamefont{Nemec}},
  \bibinfo{author}{\bibfnamefont{F.}~\bibnamefont{Lazarevic}},
  \bibinfo{author}{\bibfnamefont{P.}~\bibnamefont{Rinke}},
  \bibinfo{author}{\bibfnamefont{M.}~\bibnamefont{Scheffler}},
  \bibnamefont{and} \bibinfo{author}{\bibfnamefont{V.}~\bibnamefont{Blum}},
  \bibinfo{journal}{Phys. Rev. B} \textbf{\bibinfo{volume}{91}}
  (\bibinfo{year}{2015}), ISSN \bibinfo{issn}{1098-0121}.

\bibitem[{\citenamefont{Inoue et~al.}(2012)\citenamefont{Inoue, Kageshima,
  Kangawa, and Kakimoto}}]{InouePRB2012}
\bibinfo{author}{\bibfnamefont{M.}~\bibnamefont{Inoue}},
  \bibinfo{author}{\bibfnamefont{H.}~\bibnamefont{Kageshima}},
  \bibinfo{author}{\bibfnamefont{Y.}~\bibnamefont{Kangawa}}, \bibnamefont{and}
  \bibinfo{author}{\bibfnamefont{K.}~\bibnamefont{Kakimoto}},
  \bibinfo{journal}{Phys. Rev. B} \textbf{\bibinfo{volume}{86}}
  (\bibinfo{year}{2012}), ISSN \bibinfo{issn}{1098-0121}.

\bibitem[{\citenamefont{Kageshima et~al.}(2013)\citenamefont{Kageshima, Hibino,
  Yamaguchi, and Nagase}}]{KageshimaPRB2013}
\bibinfo{author}{\bibfnamefont{H.}~\bibnamefont{Kageshima}},
  \bibinfo{author}{\bibfnamefont{H.}~\bibnamefont{Hibino}},
  \bibinfo{author}{\bibfnamefont{H.}~\bibnamefont{Yamaguchi}},
  \bibnamefont{and} \bibinfo{author}{\bibfnamefont{M.}~\bibnamefont{Nagase}},
  \bibinfo{journal}{Phys. Rev. B} \textbf{\bibinfo{volume}{88}}
  (\bibinfo{year}{2013}), ISSN \bibinfo{issn}{1098-0121}.

\bibitem[{\citenamefont{Borysiuk et~al.}(2012)\citenamefont{Borysiuk, Soltys,
  Bozek, Piechota, Krukowski, Strupinski, Baranowski, and
  Stepniewski}}]{BorysiukPRB2012}
\bibinfo{author}{\bibfnamefont{J.}~\bibnamefont{Borysiuk}},
  \bibinfo{author}{\bibfnamefont{J.}~\bibnamefont{Soltys}},
  \bibinfo{author}{\bibfnamefont{R.}~\bibnamefont{Bozek}},
  \bibinfo{author}{\bibfnamefont{J.}~\bibnamefont{Piechota}},
  \bibinfo{author}{\bibfnamefont{S.}~\bibnamefont{Krukowski}},
  \bibinfo{author}{\bibfnamefont{W.}~\bibnamefont{Strupinski}},
  \bibinfo{author}{\bibfnamefont{J.~M.} \bibnamefont{Baranowski}},
  \bibnamefont{and}
  \bibinfo{author}{\bibfnamefont{R.}~\bibnamefont{Stepniewski}},
  \bibinfo{journal}{Phys. Rev. B} \textbf{\bibinfo{volume}{85}}
  (\bibinfo{year}{2012}), ISSN \bibinfo{issn}{1098-0121}.

\bibitem[{\citenamefont{Sun et~al.}(2011)\citenamefont{Sun, Liu, Rhim, Jia,
  Xue, Weinert, and Li}}]{SunPRB2011}
\bibinfo{author}{\bibfnamefont{G.~F.} \bibnamefont{Sun}},
  \bibinfo{author}{\bibfnamefont{Y.}~\bibnamefont{Liu}},
  \bibinfo{author}{\bibfnamefont{S.~H.} \bibnamefont{Rhim}},
  \bibinfo{author}{\bibfnamefont{J.~F.} \bibnamefont{Jia}},
  \bibinfo{author}{\bibfnamefont{Q.~K.} \bibnamefont{Xue}},
  \bibinfo{author}{\bibfnamefont{M.}~\bibnamefont{Weinert}}, \bibnamefont{and}
  \bibinfo{author}{\bibfnamefont{L.}~\bibnamefont{Li}}, \bibinfo{journal}{Phys.
  Rev. B} \textbf{\bibinfo{volume}{84}} (\bibinfo{year}{2011}), ISSN
  \bibinfo{issn}{1098-0121}.

\bibitem[{\citenamefont{Ostler et~al.}(2013)\citenamefont{Ostler, Deretzis,
  Mammadov, Giannazzo, Nicotra, Spinella, Seyller, and
  La~Magna}}]{OstlerPRB2013}
\bibinfo{author}{\bibfnamefont{M.}~\bibnamefont{Ostler}},
  \bibinfo{author}{\bibfnamefont{I.}~\bibnamefont{Deretzis}},
  \bibinfo{author}{\bibfnamefont{S.}~\bibnamefont{Mammadov}},
  \bibinfo{author}{\bibfnamefont{F.}~\bibnamefont{Giannazzo}},
  \bibinfo{author}{\bibfnamefont{G.}~\bibnamefont{Nicotra}},
  \bibinfo{author}{\bibfnamefont{C.}~\bibnamefont{Spinella}},
  \bibinfo{author}{\bibfnamefont{T.}~\bibnamefont{Seyller}}, \bibnamefont{and}
  \bibinfo{author}{\bibfnamefont{A.}~\bibnamefont{La~Magna}},
  \bibinfo{journal}{Phys. Rev. B} \textbf{\bibinfo{volume}{88}}
  (\bibinfo{year}{2013}), ISSN \bibinfo{issn}{1098-0121}.

\bibitem[{\citenamefont{Ohta et~al.}(2010)\citenamefont{Ohta, Bartelt, Nie,
  Thuermer, and Kellogg}}]{OhtaPRB2010}
\bibinfo{author}{\bibfnamefont{T.}~\bibnamefont{Ohta}},
  \bibinfo{author}{\bibfnamefont{N.~C.} \bibnamefont{Bartelt}},
  \bibinfo{author}{\bibfnamefont{S.}~\bibnamefont{Nie}},
  \bibinfo{author}{\bibfnamefont{K.}~\bibnamefont{Thuermer}}, \bibnamefont{and}
  \bibinfo{author}{\bibfnamefont{G.~L.} \bibnamefont{Kellogg}},
  \bibinfo{journal}{Phys. Rev. B} \textbf{\bibinfo{volume}{81}}
  (\bibinfo{year}{2010}), ISSN \bibinfo{issn}{1098-0121}.

\bibitem[{\citenamefont{Bolen et~al.}(2010)\citenamefont{Bolen, Shen, Gu,
  Colby, Stach, Ye, and Capano}}]{BolenJEM2010}
\bibinfo{author}{\bibfnamefont{M.~L.} \bibnamefont{Bolen}},
  \bibinfo{author}{\bibfnamefont{T.}~\bibnamefont{Shen}},
  \bibinfo{author}{\bibfnamefont{J.~J.} \bibnamefont{Gu}},
  \bibinfo{author}{\bibfnamefont{R.}~\bibnamefont{Colby}},
  \bibinfo{author}{\bibfnamefont{E.~A.} \bibnamefont{Stach}},
  \bibinfo{author}{\bibfnamefont{P.~D.} \bibnamefont{Ye}}, \bibnamefont{and}
  \bibinfo{author}{\bibfnamefont{M.~A.} \bibnamefont{Capano}},
  \bibinfo{journal}{Journal of Electronic Materials}
  \textbf{\bibinfo{volume}{39}}, \bibinfo{pages}{2696} (\bibinfo{year}{2010}),
  ISSN \bibinfo{issn}{0361-5235}, \bibinfo{note}{139th TMS Annual Meeting and
  Exhibition on Pb-Free Solders and Emerging Interconnect and Packaging
  Technologies, Seattle, WA, FEB 14-18, 2010}.

\bibitem[{\citenamefont{Kruskopf et~al.}(2016)\citenamefont{Kruskopf, Pakdehi,
  Pierz, Wundrack, Stosch, Dziomba, Goetz, Baringhaus, Aprojanz, Tegenkamp
  et~al.}}]{Kruskopf2DMaterials2016}
\bibinfo{author}{\bibfnamefont{M.}~\bibnamefont{Kruskopf}},
  \bibinfo{author}{\bibfnamefont{D.~M.} \bibnamefont{Pakdehi}},
  \bibinfo{author}{\bibfnamefont{K.}~\bibnamefont{Pierz}},
  \bibinfo{author}{\bibfnamefont{S.}~\bibnamefont{Wundrack}},
  \bibinfo{author}{\bibfnamefont{R.}~\bibnamefont{Stosch}},
  \bibinfo{author}{\bibfnamefont{T.}~\bibnamefont{Dziomba}},
  \bibinfo{author}{\bibfnamefont{M.}~\bibnamefont{Goetz}},
  \bibinfo{author}{\bibfnamefont{J.}~\bibnamefont{Baringhaus}},
  \bibinfo{author}{\bibfnamefont{J.}~\bibnamefont{Aprojanz}},
  \bibinfo{author}{\bibfnamefont{C.}~\bibnamefont{Tegenkamp}},
  \bibnamefont{et~al.}, \bibinfo{journal}{2D Materials}
  \textbf{\bibinfo{volume}{3}} (\bibinfo{year}{2016}), ISSN
  \bibinfo{issn}{2053-1583}.

\bibitem[{\citenamefont{Palmer et~al.}(2014)\citenamefont{Palmer, Kunc, Hu,
  Hankinson, Guo, Berger, and de~Heer}}]{PalmerAPL2014}
\bibinfo{author}{\bibfnamefont{J.}~\bibnamefont{Palmer}},
  \bibinfo{author}{\bibfnamefont{J.}~\bibnamefont{Kunc}},
  \bibinfo{author}{\bibfnamefont{Y.}~\bibnamefont{Hu}},
  \bibinfo{author}{\bibfnamefont{J.}~\bibnamefont{Hankinson}},
  \bibinfo{author}{\bibfnamefont{Z.}~\bibnamefont{Guo}},
  \bibinfo{author}{\bibfnamefont{C.}~\bibnamefont{Berger}}, \bibnamefont{and}
  \bibinfo{author}{\bibfnamefont{W.~A.} \bibnamefont{de~Heer}},
  \bibinfo{journal}{Appl. Phys. Lett.} \textbf{\bibinfo{volume}{105}}
  (\bibinfo{year}{2014}), ISSN \bibinfo{issn}{0003-6951}.

\bibitem[{\citenamefont{Hass et~al.}(2008)\citenamefont{Hass, de~Heer, and
  Conrad}}]{HassDeHeer2008}
\bibinfo{author}{\bibfnamefont{J.}~\bibnamefont{Hass}},
  \bibinfo{author}{\bibfnamefont{W.~A.} \bibnamefont{de~Heer}},
  \bibnamefont{and} \bibinfo{author}{\bibfnamefont{E.~H.}
  \bibnamefont{Conrad}}, \bibinfo{journal}{Jour. Phys.-Cond. Mat.}
  \textbf{\bibinfo{volume}{20}} (\bibinfo{year}{2008}), ISSN
  \bibinfo{issn}{0953-8984}.

\bibitem[{\citenamefont{Lu et~al.}(2012)\citenamefont{Lu, Barbosa, Clarke,
  Eyink, Grazulis, Mitchel, and Boeckl}}]{LuWeijieJPC2012}
\bibinfo{author}{\bibfnamefont{W.}~\bibnamefont{Lu}},
  \bibinfo{author}{\bibfnamefont{R.}~\bibnamefont{Barbosa}},
  \bibinfo{author}{\bibfnamefont{E.}~\bibnamefont{Clarke}},
  \bibinfo{author}{\bibfnamefont{K.}~\bibnamefont{Eyink}},
  \bibinfo{author}{\bibfnamefont{L.}~\bibnamefont{Grazulis}},
  \bibinfo{author}{\bibfnamefont{W.~C.} \bibnamefont{Mitchel}},
  \bibnamefont{and} \bibinfo{author}{\bibfnamefont{J.~J.}
  \bibnamefont{Boeckl}}, \bibinfo{journal}{Jour. Phys. Chem. C}
  \textbf{\bibinfo{volume}{116}}, \bibinfo{pages}{15342}
  (\bibinfo{year}{2012}), ISSN \bibinfo{issn}{1932-7447}.

\bibitem[{\citenamefont{Lu et~al.}(2010)\citenamefont{Lu, Boeckl, and
  Mitchel}}]{WeijieJoPD2010}
\bibinfo{author}{\bibfnamefont{W.}~\bibnamefont{Lu}},
  \bibinfo{author}{\bibfnamefont{J.~J.} \bibnamefont{Boeckl}},
  \bibnamefont{and} \bibinfo{author}{\bibfnamefont{W.~C.}
  \bibnamefont{Mitchel}}, \bibinfo{journal}{Jour. Phys. D-Appl. Phys.}
  \textbf{\bibinfo{volume}{43}} (\bibinfo{year}{2010}), ISSN
  \bibinfo{issn}{0022-3727}.

\bibitem[{\citenamefont{Kwak et~al.}(2013)\citenamefont{Kwak, Kwon, Chu, Choi,
  Lee, Kim, Shin, Park, Park, and Kwon}}]{KwakPCCP2013}
\bibinfo{author}{\bibfnamefont{J.}~\bibnamefont{Kwak}},
  \bibinfo{author}{\bibfnamefont{T.-Y.} \bibnamefont{Kwon}},
  \bibinfo{author}{\bibfnamefont{J.~H.} \bibnamefont{Chu}},
  \bibinfo{author}{\bibfnamefont{J.-K.} \bibnamefont{Choi}},
  \bibinfo{author}{\bibfnamefont{M.-S.} \bibnamefont{Lee}},
  \bibinfo{author}{\bibfnamefont{S.~Y.} \bibnamefont{Kim}},
  \bibinfo{author}{\bibfnamefont{H.-J.} \bibnamefont{Shin}},
  \bibinfo{author}{\bibfnamefont{K.}~\bibnamefont{Park}},
  \bibinfo{author}{\bibfnamefont{J.-U.} \bibnamefont{Park}}, \bibnamefont{and}
  \bibinfo{author}{\bibfnamefont{S.-Y.} \bibnamefont{Kwon}},
  \bibinfo{journal}{Phys. Chem. Chem. Phys.} \textbf{\bibinfo{volume}{15}},
  \bibinfo{pages}{10446} (\bibinfo{year}{2013}), ISSN
  \bibinfo{issn}{1463-9076}.

\bibitem[{\citenamefont{Pattinson et~al.}(2012)\citenamefont{Pattinson,
  Ranganathan, Murakami, Koziol, and Windle}}]{PattinsonACSNano2012}
\bibinfo{author}{\bibfnamefont{S.~W.} \bibnamefont{Pattinson}},
  \bibinfo{author}{\bibfnamefont{V.}~\bibnamefont{Ranganathan}},
  \bibinfo{author}{\bibfnamefont{H.~K.} \bibnamefont{Murakami}},
  \bibinfo{author}{\bibfnamefont{K.~K.~K.} \bibnamefont{Koziol}},
  \bibnamefont{and} \bibinfo{author}{\bibfnamefont{A.~H.}
  \bibnamefont{Windle}}, \bibinfo{journal}{ACS Nano}
  \textbf{\bibinfo{volume}{6}}, \bibinfo{pages}{7723} (\bibinfo{year}{2012}),
  ISSN \bibinfo{issn}{1936-0851}.

\bibitem[{\citenamefont{Norimatsu et~al.}(2011)\citenamefont{Norimatsu, Takada,
  and Kusunoki}}]{NorimatsuPRB2011}
\bibinfo{author}{\bibfnamefont{W.}~\bibnamefont{Norimatsu}},
  \bibinfo{author}{\bibfnamefont{J.}~\bibnamefont{Takada}}, \bibnamefont{and}
  \bibinfo{author}{\bibfnamefont{M.}~\bibnamefont{Kusunoki}},
  \bibinfo{journal}{Phys. Rev. B} \textbf{\bibinfo{volume}{84}}
  (\bibinfo{year}{2011}), ISSN \bibinfo{issn}{1098-0121}.

\bibitem[{\citenamefont{de~Heer et~al.}(2011)\citenamefont{de~Heer, Berger,
  Ruan, Sprinkle, Li, Hu, Zhang, Hankinson, and Conrad}}]{deHeerPNAS2011}
\bibinfo{author}{\bibfnamefont{W.~A.} \bibnamefont{de~Heer}},
  \bibinfo{author}{\bibfnamefont{C.}~\bibnamefont{Berger}},
  \bibinfo{author}{\bibfnamefont{M.}~\bibnamefont{Ruan}},
  \bibinfo{author}{\bibfnamefont{M.}~\bibnamefont{Sprinkle}},
  \bibinfo{author}{\bibfnamefont{X.}~\bibnamefont{Li}},
  \bibinfo{author}{\bibfnamefont{Y.}~\bibnamefont{Hu}},
  \bibinfo{author}{\bibfnamefont{B.}~\bibnamefont{Zhang}},
  \bibinfo{author}{\bibfnamefont{J.}~\bibnamefont{Hankinson}},
  \bibnamefont{and} \bibinfo{author}{\bibfnamefont{E.}~\bibnamefont{Conrad}},
  \bibinfo{journal}{Proceedings of the National Academy of Sciences of the
  United States of America} \textbf{\bibinfo{volume}{108}},
  \bibinfo{pages}{16900} (\bibinfo{year}{2011}), ISSN
  \bibinfo{issn}{0027-8424}.

\bibitem[{\citenamefont{Schmidt et~al.}(2011)\citenamefont{Schmidt, Ohta, and
  Beechem}}]{SchmidtPRB2011}
\bibinfo{author}{\bibfnamefont{D.~A.} \bibnamefont{Schmidt}},
  \bibinfo{author}{\bibfnamefont{T.}~\bibnamefont{Ohta}}, \bibnamefont{and}
  \bibinfo{author}{\bibfnamefont{T.~E.} \bibnamefont{Beechem}},
  \bibinfo{journal}{Phys. Rev. B} \textbf{\bibinfo{volume}{84}}
  (\bibinfo{year}{2011}), ISSN \bibinfo{issn}{1098-0121}.

\bibitem[{\citenamefont{Stampfer et~al.}(2007)\citenamefont{Stampfer, Molitor,
  Graf, Ensslin, Jungen, Hierold, and Wirtz}}]{StampferAPL2007}
\bibinfo{author}{\bibfnamefont{C.}~\bibnamefont{Stampfer}},
  \bibinfo{author}{\bibfnamefont{F.}~\bibnamefont{Molitor}},
  \bibinfo{author}{\bibfnamefont{D.}~\bibnamefont{Graf}},
  \bibinfo{author}{\bibfnamefont{K.}~\bibnamefont{Ensslin}},
  \bibinfo{author}{\bibfnamefont{A.}~\bibnamefont{Jungen}},
  \bibinfo{author}{\bibfnamefont{C.}~\bibnamefont{Hierold}}, \bibnamefont{and}
  \bibinfo{author}{\bibfnamefont{L.}~\bibnamefont{Wirtz}},
  \bibinfo{journal}{Appl. Phys. Lett.} \textbf{\bibinfo{volume}{91}}
  (\bibinfo{year}{2007}), ISSN \bibinfo{issn}{0003-6951}.

\bibitem[{\citenamefont{Lazzeri and Mauri}(2006)}]{LazzeriPRL2006}
\bibinfo{author}{\bibfnamefont{M.}~\bibnamefont{Lazzeri}} \bibnamefont{and}
  \bibinfo{author}{\bibfnamefont{F.}~\bibnamefont{Mauri}},
  \bibinfo{journal}{Phys. Rev. Lett.} \textbf{\bibinfo{volume}{97}}
  (\bibinfo{year}{2006}), ISSN \bibinfo{issn}{0031-9007}.

\bibitem[{\citenamefont{Das et~al.}(2008)\citenamefont{Das, Pisana,
  Chakraborty, Piscanec, Saha, Waghmare, Novoselov, Krishnamurthy, Geim,
  Ferrari et~al.}}]{DasNatureNanotech2008}
\bibinfo{author}{\bibfnamefont{A.}~\bibnamefont{Das}},
  \bibinfo{author}{\bibfnamefont{S.}~\bibnamefont{Pisana}},
  \bibinfo{author}{\bibfnamefont{B.}~\bibnamefont{Chakraborty}},
  \bibinfo{author}{\bibfnamefont{S.}~\bibnamefont{Piscanec}},
  \bibinfo{author}{\bibfnamefont{S.~K.} \bibnamefont{Saha}},
  \bibinfo{author}{\bibfnamefont{U.~V.} \bibnamefont{Waghmare}},
  \bibinfo{author}{\bibfnamefont{K.~S.} \bibnamefont{Novoselov}},
  \bibinfo{author}{\bibfnamefont{H.~R.} \bibnamefont{Krishnamurthy}},
  \bibinfo{author}{\bibfnamefont{A.~K.} \bibnamefont{Geim}},
  \bibinfo{author}{\bibfnamefont{A.~C.} \bibnamefont{Ferrari}},
  \bibnamefont{et~al.}, \bibinfo{journal}{Nature Nanotech.}
  \textbf{\bibinfo{volume}{3}}, \bibinfo{pages}{210} (\bibinfo{year}{2008}),
  ISSN \bibinfo{issn}{1748-3387}.

\bibitem[{\citenamefont{Dresselhaus et~al.}(2010)\citenamefont{Dresselhaus,
  Jorio, Souza~Filho, and Saito}}]{Dresselhaus2010}
\bibinfo{author}{\bibfnamefont{M.~S.} \bibnamefont{Dresselhaus}},
  \bibinfo{author}{\bibfnamefont{A.}~\bibnamefont{Jorio}},
  \bibinfo{author}{\bibfnamefont{A.~G.} \bibnamefont{Souza~Filho}},
  \bibnamefont{and} \bibinfo{author}{\bibfnamefont{R.}~\bibnamefont{Saito}},
  \bibinfo{journal}{Philosophical Transactions of the Royal Society
  A-Mathematical Physical and Engineering Sciences}
  \textbf{\bibinfo{volume}{368}}, \bibinfo{pages}{5355} (\bibinfo{year}{2010}),
  ISSN \bibinfo{issn}{1364-503X}.

\bibitem[{\citenamefont{Cancado et~al.}(2011)\citenamefont{Cancado, Jorio,
  Martins~Ferreira, Stavale, Achete, Capaz, Moutinho, Lombardo, Kulmala, and
  Ferrari}}]{CancadoNanoLetters2011}
\bibinfo{author}{\bibfnamefont{L.~G.} \bibnamefont{Cancado}},
  \bibinfo{author}{\bibfnamefont{A.}~\bibnamefont{Jorio}},
  \bibinfo{author}{\bibfnamefont{E.~H.} \bibnamefont{Martins~Ferreira}},
  \bibinfo{author}{\bibfnamefont{F.}~\bibnamefont{Stavale}},
  \bibinfo{author}{\bibfnamefont{C.~A.} \bibnamefont{Achete}},
  \bibinfo{author}{\bibfnamefont{R.~B.} \bibnamefont{Capaz}},
  \bibinfo{author}{\bibfnamefont{M.~V.~O.} \bibnamefont{Moutinho}},
  \bibinfo{author}{\bibfnamefont{A.}~\bibnamefont{Lombardo}},
  \bibinfo{author}{\bibfnamefont{T.~S.} \bibnamefont{Kulmala}},
  \bibnamefont{and} \bibinfo{author}{\bibfnamefont{A.~C.}
  \bibnamefont{Ferrari}}, \bibinfo{journal}{Nano Lett.}
  \textbf{\bibinfo{volume}{11}}, \bibinfo{pages}{3190} (\bibinfo{year}{2011}),
  ISSN \bibinfo{issn}{1530-6984}.

\bibitem[{\citenamefont{Cancado et~al.}(2006)\citenamefont{Cancado, Takai,
  Enoki, Endo, Kim, Mizusaki, Jorio, Coelho, Magalhaes-Paniago, and
  Pimenta}}]{CancadoAPL2006}
\bibinfo{author}{\bibfnamefont{L.}~\bibnamefont{Cancado}},
  \bibinfo{author}{\bibfnamefont{K.}~\bibnamefont{Takai}},
  \bibinfo{author}{\bibfnamefont{T.}~\bibnamefont{Enoki}},
  \bibinfo{author}{\bibfnamefont{M.}~\bibnamefont{Endo}},
  \bibinfo{author}{\bibfnamefont{Y.}~\bibnamefont{Kim}},
  \bibinfo{author}{\bibfnamefont{H.}~\bibnamefont{Mizusaki}},
  \bibinfo{author}{\bibfnamefont{A.}~\bibnamefont{Jorio}},
  \bibinfo{author}{\bibfnamefont{L.}~\bibnamefont{Coelho}},
  \bibinfo{author}{\bibfnamefont{R.}~\bibnamefont{Magalhaes-Paniago}},
  \bibnamefont{and} \bibinfo{author}{\bibfnamefont{M.}~\bibnamefont{Pimenta}},
  \bibinfo{journal}{Appl. Phys. Lett.} \textbf{\bibinfo{volume}{88}}
  (\bibinfo{year}{2006}), ISSN \bibinfo{issn}{0003-6951}.

\bibitem[{\citenamefont{Ferrari and Basko}(2013)}]{FerrariNatureNanotech2013}
\bibinfo{author}{\bibfnamefont{A.~C.} \bibnamefont{Ferrari}} \bibnamefont{and}
  \bibinfo{author}{\bibfnamefont{D.~M.} \bibnamefont{Basko}},
  \bibinfo{journal}{Nature Nanotech.} \textbf{\bibinfo{volume}{8}},
  \bibinfo{pages}{235} (\bibinfo{year}{2013}), ISSN \bibinfo{issn}{1748-3387}.

\bibitem[{\citenamefont{Hwang et~al.}(2013)\citenamefont{Hwang, Lin, Hwang,
  Chang, Chattopadhyay, Chen, Chen, Chiang, Tsai, Chen
  et~al.}}]{HwangNanotech2013}
\bibinfo{author}{\bibfnamefont{J.-S.} \bibnamefont{Hwang}},
  \bibinfo{author}{\bibfnamefont{Y.-H.} \bibnamefont{Lin}},
  \bibinfo{author}{\bibfnamefont{J.-Y.} \bibnamefont{Hwang}},
  \bibinfo{author}{\bibfnamefont{R.}~\bibnamefont{Chang}},
  \bibinfo{author}{\bibfnamefont{S.}~\bibnamefont{Chattopadhyay}},
  \bibinfo{author}{\bibfnamefont{C.-J.} \bibnamefont{Chen}},
  \bibinfo{author}{\bibfnamefont{P.}~\bibnamefont{Chen}},
  \bibinfo{author}{\bibfnamefont{H.-P.} \bibnamefont{Chiang}},
  \bibinfo{author}{\bibfnamefont{T.-R.} \bibnamefont{Tsai}},
  \bibinfo{author}{\bibfnamefont{L.-C.} \bibnamefont{Chen}},
  \bibnamefont{et~al.}, \bibinfo{journal}{Nanotech.}
  \textbf{\bibinfo{volume}{24}} (\bibinfo{year}{2013}), ISSN
  \bibinfo{issn}{0957-4484}.

\bibitem[{\citenamefont{Lee et~al.}(2008{\natexlab{a}})\citenamefont{Lee,
  Riedl, Krauss, von Klitzing, Starke, and Smet}}]{LeeSmetNanoLetters2008}
\bibinfo{author}{\bibfnamefont{D.~S.} \bibnamefont{Lee}},
  \bibinfo{author}{\bibfnamefont{C.}~\bibnamefont{Riedl}},
  \bibinfo{author}{\bibfnamefont{B.}~\bibnamefont{Krauss}},
  \bibinfo{author}{\bibfnamefont{K.}~\bibnamefont{von Klitzing}},
  \bibinfo{author}{\bibfnamefont{U.}~\bibnamefont{Starke}}, \bibnamefont{and}
  \bibinfo{author}{\bibfnamefont{J.~H.} \bibnamefont{Smet}},
  \bibinfo{journal}{Nano Lett.} \textbf{\bibinfo{volume}{8}},
  \bibinfo{pages}{4320} (\bibinfo{year}{2008}{\natexlab{a}}), ISSN
  \bibinfo{issn}{1530-6984}.

\bibitem[{\citenamefont{Robinson et~al.}(2009)\citenamefont{Robinson,
  Wetherington, Tedesco, Campbell, Weng, Stitt, Fanton, Frantz, Snyder, VanMil
  et~al.}}]{RobinsonNanoLett2009}
\bibinfo{author}{\bibfnamefont{J.~A.} \bibnamefont{Robinson}},
  \bibinfo{author}{\bibfnamefont{M.}~\bibnamefont{Wetherington}},
  \bibinfo{author}{\bibfnamefont{J.~L.} \bibnamefont{Tedesco}},
  \bibinfo{author}{\bibfnamefont{P.~M.} \bibnamefont{Campbell}},
  \bibinfo{author}{\bibfnamefont{X.}~\bibnamefont{Weng}},
  \bibinfo{author}{\bibfnamefont{J.}~\bibnamefont{Stitt}},
  \bibinfo{author}{\bibfnamefont{M.~A.} \bibnamefont{Fanton}},
  \bibinfo{author}{\bibfnamefont{E.}~\bibnamefont{Frantz}},
  \bibinfo{author}{\bibfnamefont{D.}~\bibnamefont{Snyder}},
  \bibinfo{author}{\bibfnamefont{B.~L.} \bibnamefont{VanMil}},
  \bibnamefont{et~al.}, \bibinfo{journal}{Nano Lett.}
  \textbf{\bibinfo{volume}{9}}, \bibinfo{pages}{2873} (\bibinfo{year}{2009}),
  ISSN \bibinfo{issn}{1530-6984}.

\bibitem[{\citenamefont{Pozzo et~al.}(2011)\citenamefont{Pozzo, Alfe, Lacovig,
  Hofmann, Lizzit, and Baraldi}}]{PozzoPRL2011}
\bibinfo{author}{\bibfnamefont{M.}~\bibnamefont{Pozzo}},
  \bibinfo{author}{\bibfnamefont{D.}~\bibnamefont{Alfe}},
  \bibinfo{author}{\bibfnamefont{P.}~\bibnamefont{Lacovig}},
  \bibinfo{author}{\bibfnamefont{P.}~\bibnamefont{Hofmann}},
  \bibinfo{author}{\bibfnamefont{S.}~\bibnamefont{Lizzit}}, \bibnamefont{and}
  \bibinfo{author}{\bibfnamefont{A.}~\bibnamefont{Baraldi}},
  \bibinfo{journal}{Phys. Rev. Lett.} \textbf{\bibinfo{volume}{106}}
  (\bibinfo{year}{2011}), ISSN \bibinfo{issn}{0031-9007}.

\bibitem[{\citenamefont{Emtsev et~al.}(2008)\citenamefont{Emtsev, Speck,
  Seyller, Ley, and Riley}}]{EmtsevPRB2008}
\bibinfo{author}{\bibfnamefont{K.~V.} \bibnamefont{Emtsev}},
  \bibinfo{author}{\bibfnamefont{F.}~\bibnamefont{Speck}},
  \bibinfo{author}{\bibfnamefont{T.}~\bibnamefont{Seyller}},
  \bibinfo{author}{\bibfnamefont{L.}~\bibnamefont{Ley}}, \bibnamefont{and}
  \bibinfo{author}{\bibfnamefont{J.~D.} \bibnamefont{Riley}},
  \bibinfo{journal}{Phys. Rev. B} \textbf{\bibinfo{volume}{77}}
  (\bibinfo{year}{2008}), ISSN \bibinfo{issn}{1098-0121}.

\bibitem[{\citenamefont{Partovi-Azar et~al.}(2011)\citenamefont{Partovi-Azar,
  Nafari, and Tabar}}]{AzarPRB2011}
\bibinfo{author}{\bibfnamefont{P.}~\bibnamefont{Partovi-Azar}},
  \bibinfo{author}{\bibfnamefont{N.}~\bibnamefont{Nafari}}, \bibnamefont{and}
  \bibinfo{author}{\bibfnamefont{M.~R.~R.} \bibnamefont{Tabar}},
  \bibinfo{journal}{Phys. Rev. B} \textbf{\bibinfo{volume}{83}}
  (\bibinfo{year}{2011}), ISSN \bibinfo{issn}{1098-0121}.

\bibitem[{\citenamefont{Yan et~al.}(2007)\citenamefont{Yan, Zhang, Kim, and
  Pinczuk}}]{YanPRL2007}
\bibinfo{author}{\bibfnamefont{J.}~\bibnamefont{Yan}},
  \bibinfo{author}{\bibfnamefont{Y.}~\bibnamefont{Zhang}},
  \bibinfo{author}{\bibfnamefont{P.}~\bibnamefont{Kim}}, \bibnamefont{and}
  \bibinfo{author}{\bibfnamefont{A.}~\bibnamefont{Pinczuk}},
  \bibinfo{journal}{Phys. Rev. Lett.} \textbf{\bibinfo{volume}{98}}
  (\bibinfo{year}{2007}), ISSN \bibinfo{issn}{0031-9007}.

\bibitem[{\citenamefont{Tedesco et~al.}(2009)\citenamefont{Tedesco, VanMil,
  Myers-Ward, McCrate, Kitt, Campbell, Jernigan, Culbertson, Eddy, and
  Gaskill}}]{TedescoGaskill2009}
\bibinfo{author}{\bibfnamefont{J.~L.} \bibnamefont{Tedesco}},
  \bibinfo{author}{\bibfnamefont{B.~L.} \bibnamefont{VanMil}},
  \bibinfo{author}{\bibfnamefont{R.~L.} \bibnamefont{Myers-Ward}},
  \bibinfo{author}{\bibfnamefont{J.~M.} \bibnamefont{McCrate}},
  \bibinfo{author}{\bibfnamefont{S.~A.} \bibnamefont{Kitt}},
  \bibinfo{author}{\bibfnamefont{P.~M.} \bibnamefont{Campbell}},
  \bibinfo{author}{\bibfnamefont{G.~G.} \bibnamefont{Jernigan}},
  \bibinfo{author}{\bibfnamefont{J.~C.} \bibnamefont{Culbertson}},
  \bibinfo{author}{\bibfnamefont{C.~R.} \bibnamefont{Eddy},
  \bibfnamefont{Jr.}}, \bibnamefont{and} \bibinfo{author}{\bibfnamefont{D.~K.}
  \bibnamefont{Gaskill}}, \bibinfo{journal}{Appl. Phys. Lett.}
  \textbf{\bibinfo{volume}{95}} (\bibinfo{year}{2009}), ISSN
  \bibinfo{issn}{0003-6951}.

\bibitem[{\citenamefont{Curtin et~al.}(2011)\citenamefont{Curtin, Fuhrer,
  Tedesco, Myers-Ward, Eddy, and Gaskill}}]{CurtinTedescoGaskillAPL2011}
\bibinfo{author}{\bibfnamefont{A.~E.} \bibnamefont{Curtin}},
  \bibinfo{author}{\bibfnamefont{M.~S.} \bibnamefont{Fuhrer}},
  \bibinfo{author}{\bibfnamefont{J.~L.} \bibnamefont{Tedesco}},
  \bibinfo{author}{\bibfnamefont{R.~L.} \bibnamefont{Myers-Ward}},
  \bibinfo{author}{\bibfnamefont{C.~R.} \bibnamefont{Eddy},
  \bibfnamefont{Jr.}}, \bibnamefont{and} \bibinfo{author}{\bibfnamefont{D.~K.}
  \bibnamefont{Gaskill}}, \bibinfo{journal}{Appl. Phys. Lett.}
  \textbf{\bibinfo{volume}{98}} (\bibinfo{year}{2011}), ISSN
  \bibinfo{issn}{0003-6951}.

\bibitem[{\citenamefont{Wiedmann et~al.}(2011)\citenamefont{Wiedmann, van
  Elferen, Kurganova, Katsnelson, Giesbers, Veligura, van Wees, Gorbachev,
  Novoselov, Maan et~al.}}]{WiedmannPRB2011}
\bibinfo{author}{\bibfnamefont{S.}~\bibnamefont{Wiedmann}},
  \bibinfo{author}{\bibfnamefont{H.~J.} \bibnamefont{van Elferen}},
  \bibinfo{author}{\bibfnamefont{E.~V.} \bibnamefont{Kurganova}},
  \bibinfo{author}{\bibfnamefont{M.~I.} \bibnamefont{Katsnelson}},
  \bibinfo{author}{\bibfnamefont{A.~J.~M.} \bibnamefont{Giesbers}},
  \bibinfo{author}{\bibfnamefont{A.}~\bibnamefont{Veligura}},
  \bibinfo{author}{\bibfnamefont{B.~J.} \bibnamefont{van Wees}},
  \bibinfo{author}{\bibfnamefont{R.~V.} \bibnamefont{Gorbachev}},
  \bibinfo{author}{\bibfnamefont{K.~S.} \bibnamefont{Novoselov}},
  \bibinfo{author}{\bibfnamefont{J.~C.} \bibnamefont{Maan}},
  \bibnamefont{et~al.}, \bibinfo{journal}{Phys. Rev. B}
  \textbf{\bibinfo{volume}{84}} (\bibinfo{year}{2011}), ISSN
  \bibinfo{issn}{1098-0121}.

\bibitem[{\citenamefont{Huang et~al.}(2015)\citenamefont{Huang,
  Alexander-Webber, Baker, Janssen, Tzalenchuk, Antonov, Yager, Lara-Avila,
  Kubatkin, Yakimova et~al.}}]{HuangPRB2015}
\bibinfo{author}{\bibfnamefont{J.}~\bibnamefont{Huang}},
  \bibinfo{author}{\bibfnamefont{J.~A.} \bibnamefont{Alexander-Webber}},
  \bibinfo{author}{\bibfnamefont{A.~M.~R.} \bibnamefont{Baker}},
  \bibinfo{author}{\bibfnamefont{T.~J. B.~M.} \bibnamefont{Janssen}},
  \bibinfo{author}{\bibfnamefont{A.}~\bibnamefont{Tzalenchuk}},
  \bibinfo{author}{\bibfnamefont{V.}~\bibnamefont{Antonov}},
  \bibinfo{author}{\bibfnamefont{T.}~\bibnamefont{Yager}},
  \bibinfo{author}{\bibfnamefont{S.}~\bibnamefont{Lara-Avila}},
  \bibinfo{author}{\bibfnamefont{S.}~\bibnamefont{Kubatkin}},
  \bibinfo{author}{\bibfnamefont{R.}~\bibnamefont{Yakimova}},
  \bibnamefont{et~al.}, \bibinfo{journal}{Phys. Rev. B}
  \textbf{\bibinfo{volume}{92}} (\bibinfo{year}{2015}), ISSN
  \bibinfo{issn}{1098-0121}.

\bibitem[{\citenamefont{Rajput et~al.}(2014)\citenamefont{Rajput, Li, and
  Li}}]{RajputAPL2014}
\bibinfo{author}{\bibfnamefont{S.}~\bibnamefont{Rajput}},
  \bibinfo{author}{\bibfnamefont{Y.~Y.} \bibnamefont{Li}}, \bibnamefont{and}
  \bibinfo{author}{\bibfnamefont{L.}~\bibnamefont{Li}}, \bibinfo{journal}{Appl.
  Phys. Lett.S} \textbf{\bibinfo{volume}{104}} (\bibinfo{year}{2014}), ISSN
  \bibinfo{issn}{0003-6951}.

\bibitem[{\citenamefont{Martin et~al.}(2008)\citenamefont{Martin, Akerman,
  Ulbricht, Lohmann, Smet, Von~Klitzing, and Yacoby}}]{MartinNaturePhys2008}
\bibinfo{author}{\bibfnamefont{J.}~\bibnamefont{Martin}},
  \bibinfo{author}{\bibfnamefont{N.}~\bibnamefont{Akerman}},
  \bibinfo{author}{\bibfnamefont{G.}~\bibnamefont{Ulbricht}},
  \bibinfo{author}{\bibfnamefont{T.}~\bibnamefont{Lohmann}},
  \bibinfo{author}{\bibfnamefont{J.~H.} \bibnamefont{Smet}},
  \bibinfo{author}{\bibfnamefont{K.}~\bibnamefont{Von~Klitzing}},
  \bibnamefont{and} \bibinfo{author}{\bibfnamefont{A.}~\bibnamefont{Yacoby}},
  \bibinfo{journal}{Nat. Phys.} \textbf{\bibinfo{volume}{4}},
  \bibinfo{pages}{144} (\bibinfo{year}{2008}), ISSN \bibinfo{issn}{1745-2473}.

\bibitem[{\citenamefont{Shivaraman et~al.}(2009)\citenamefont{Shivaraman,
  Chandrashekhar, Boeckl, and Spencer}}]{ShivaramanJEM2009}
\bibinfo{author}{\bibfnamefont{S.}~\bibnamefont{Shivaraman}},
  \bibinfo{author}{\bibfnamefont{M.~V.~S.} \bibnamefont{Chandrashekhar}},
  \bibinfo{author}{\bibfnamefont{J.~J.} \bibnamefont{Boeckl}},
  \bibnamefont{and} \bibinfo{author}{\bibfnamefont{M.~G.}
  \bibnamefont{Spencer}}, \bibinfo{journal}{Journal of Electronic Materials}
  \textbf{\bibinfo{volume}{38}}, \bibinfo{pages}{725} (\bibinfo{year}{2009}),
  ISSN \bibinfo{issn}{0361-5235}, \bibinfo{note}{50th Electronic Materials
  Conference, Univ Calif Santa Barbara, Santa Barbara, CA, JUN, 2008}.

\bibitem[{\citenamefont{Lee et~al.}(2008{\natexlab{b}})\citenamefont{Lee,
  Riedl, Krauss, von Klitzing, Starke, and Smet}}]{LeeNanoLetters2008}
\bibinfo{author}{\bibfnamefont{D.~S.} \bibnamefont{Lee}},
  \bibinfo{author}{\bibfnamefont{C.}~\bibnamefont{Riedl}},
  \bibinfo{author}{\bibfnamefont{B.}~\bibnamefont{Krauss}},
  \bibinfo{author}{\bibfnamefont{K.}~\bibnamefont{von Klitzing}},
  \bibinfo{author}{\bibfnamefont{U.}~\bibnamefont{Starke}}, \bibnamefont{and}
  \bibinfo{author}{\bibfnamefont{J.~H.} \bibnamefont{Smet}},
  \bibinfo{journal}{Nano Lett.} \textbf{\bibinfo{volume}{8}},
  \bibinfo{pages}{4320} (\bibinfo{year}{2008}{\natexlab{b}}), ISSN
  \bibinfo{issn}{1530-6984}.

\end{thebibliography}
\end{document}